\PassOptionsToPackage{table}{xcolor}

\documentclass[10pt,journal,compsoc,dvipsnames,letterpaper]{IEEEtran}

\usepackage[mode=buildnew]{standalone}
\usepackage{amsthm}
\usepackage{newtxmath}
\usepackage{zi4}
\usepackage{amsfonts}
\usepackage{amsmath}
\usepackage{algorithmic}
\usepackage{csquotes}
\usepackage{graphicx}
\usepackage{longtable}
\usepackage{algorithm}
\usepackage{subcaption}
\usepackage{enumitem}
\usepackage{comment}
\usepackage{xspace}
\usepackage{balance}
\usepackage{fancyhdr}
\usepackage{multirow}
\usepackage{soul}
\usepackage{array}
\usepackage{booktabs}
\usepackage{caption}

\usepackage{enumitem}
\setitemize[1]{leftmargin=*}

\usepackage{siunitx}
\DeclareSIUnit{\bps}{bps}

\AtBeginEnvironment{tabular}{\footnotesize}

\usepackage{tikz}
\usepackage{pgfplots}
\usetikzlibrary{babel, positioning, arrows, arrows.meta, decorations.markings, calc, pgfplots.colorbrewer, patterns} %

\usepgfplotslibrary{patchplots,groupplots,colorbrewer}
\pgfplotsset{
	compat=newest,
	plot coordinates/math parser=false,
	every tick label/.style={font=\scriptsize},
	colormap/Paired
}

\usepackage[acronyms,nonumberlist,nopostdot,nomain,nogroupskip,acronymlists={hidden}]{glossaries}
\newglossary[algh]{hidden}{acrh}{acnh}{Hidden Acronyms}
\newacronym{3gpp}{3GPP}{3rd Generation Partnership Project}
\newacronym{4g}{4G}{4th generation}
\newacronym{5g}{5G}{5th generation}
\newacronym{6g}{6G}{6th generation}
\newacronym{5gc}{5GC}{5G Core}
\newacronym{adc}{ADC}{Analog to Digital Converter}
\newacronym{aerpaw}{AERPAW}{Aerial Experimentation and Research Platform for Advanced Wireless}
\newacronym{ai}{AI}{Artificial Intelligence}
\newacronym{aimd}{AIMD}{Additive Increase Multiplicative Decrease}
\newacronym{am}{AM}{Acknowledged Mode}
\newacronym{amc}{AMC}{Adaptive Modulation and Coding}
\newacronym{amf}{AMF}{Access and Mobility Management Function}
\newacronym{aops}{AOPS}{Adaptive Order Prediction Scheduling}
\newacronym{api}{API}{Application Programming Interface}
\newacronym{xapp}{xApp}{Intelligent Application}
\newacronym{apn}{APN}{Access Point Name}
\newacronym{ap}{AP}{Application Protocol}
\newacronym{aqm}{AQM}{Active Queue Management}
\newacronym{ausf}{AUSF}{Authentication Server Function}
\newacronym{avc}{AVC}{Advanced Video Coding}
\newacronym{awgn}{AGWN}{Additive White Gaussian Noise}
\newacronym{balia}{BALIA}{Balanced Link Adaptation Algorithm}
\newacronym{bbu}{BBU}{Base Band Unit}
\newacronym{bdp}{BDP}{Bandwidth-Delay Product}
\newacronym{ber}{BER}{Bit Error Rate}
\newacronym{bf}{BF}{Beamforming}
\newacronym{bler}{BLER}{Block Error Rate}
\newacronym{brr}{BRR}{Bayesian Ridge Regressor}
\newacronym{bs}{BS}{Base Station}
\newacronym{bsr}{BSR}{Buffer Status Report}
\newacronym{bss}{BSS}{Business Support System}
\newacronym{ca}{CA}{Carrier Aggregation}
\newacronym{caas}{CaaS}{Connectivity-as-a-Service}
\newacronym{cb}{CB}{Code Block}
\newacronym{cc}{CC}{Congestion Control}
\newacronym{ccid}{CCID}{Congestion Control ID}
\newacronym{cco}{CC}{Carrier Component}
\newacronym{cdd}{CDD}{Cyclic Delay Diversity}
\newacronym{cdf}{CDF}{Cumulative Distribution Function}
\newacronym{cdn}{CDN}{Content Distribution Network}
\newacronym{cn}{CN}{Core Network}
\newacronym{codel}{CoDel}{Controlled Delay Management}
\newacronym{comac}{COMAC}{Converged Multi-Access and Core}
\newacronym{cord}{CORD}{Central Office Re-architected as a Datacenter}
\newacronym{cornet}{CORNET}{COgnitive Radio NETwork}
\newacronym{cosmos}{COSMOS}{Cloud Enhanced Open Software Defined Mobile Wireless Testbed for City-Scale Deployment}
\newacronym{cots}{COTS}{Commercial Off-the-Shelf}
\newacronym{cp}{CP}{Control Plane}
\newacronym{cpu}{CPU}{Central Processing Unit}
\newacronym{cqi}{CQI}{Channel Quality Information}
\newacronym{cr}{CR}{Cognitive Radio}
\newacronym{cran}{CRAN}{Cloud \gls{ran}}
\newacronym{crs}{CRS}{Cell Reference Signal}
\newacronym{csi}{CSI}{Channel State Information}
\newacronym{csirs}{CSI-RS}{Channel State Information - Reference Signal}
\newacronym{cu}{CU}{Central Unit}
\newacronym{d2tcp}{D$^2$TCP}{Deadline-aware Data center TCP}
\newacronym{d3}{D$^3$}{Deadline-Driven Delivery}
\newacronym{dac}{DAC}{Digital to Analog Converter}
\newacronym{dag}{DAG}{Directed Acyclic Graph}
\newacronym{das}{DAS}{Distributed Antenna System}
\newacronym{dash}{DASH}{Dynamic Adaptive Streaming over HTTP}
\newacronym{dc}{DC}{Dual Connectivity}
\newacronym{dccp}{DCCP}{Datagram Congestion Control Protocol}
\newacronym{dce}{DCE}{Direct Code Execution}
\newacronym{dci}{DCI}{Downlink Control Information}
\newacronym{dctcp}{DCTCP}{Data Center TCP}
\newacronym{dl}{DL}{Downlink}
\newacronym{dmr}{DMR}{Deadline Miss Ratio}
\newacronym{dmrs}{DMRS}{DeModulation Reference Signal}
\newacronym{drlcc}{DRL-CC}{Deep Reinforcement Learning Congestion Control}
\newacronym{drs}{DRS}{Discovery Reference Signal}
\newacronym{du}{DU}{Distributed Unit}
\newacronym{e2e}{E2E}{end-to-end}
\newacronym{ecaas}{ECaaS}{Edge-Cloud-as-a-Service}
\newacronym{ecn}{ECN}{Explicit Congestion Notification}
\newacronym{ecdf}{ECDF}{Empirical Cumulative Density Function}
\newacronym{edf}{EDF}{Earliest Deadline First}
\newacronym{embb}{eMBB}{Enhanced Mobile Broadband}
\newacronym{empower}{EMPOWER}{EMpowering transatlantic PlatfOrms for advanced WirEless Research}
\newacronym{enb}{eNB}{evolved Node Base}
\newacronym{endc}{EN-DC}{E-UTRAN-\gls{nr} \gls{dc}}
\newacronym{epc}{EPC}{Evolved Packet Core}
\newacronym{eps}{EPS}{Evolved Packet System}
\newacronym{etsi}{ETSI}{European Telecommunications Standards Institute}
\newacronym[firstplural=Estimated Times of Arrival (ETAs)]{eta}{ETA}{Estimated Time of Arrival}
\newacronym{eutran}{E-UTRAN}{Evolved Universal Terrestrial Access Network}
\newacronym{faas}{FaaS}{Function-as-a-Service}
\newacronym{fapi}{FAPI}{Functional Application Platform Interface}
\newacronym{fdd}{FDD}{Frequency Division Duplexing}
\newacronym{fdm}{FDM}{Frequency Division Multiplexing}
\newacronym{fdma}{FDMA}{Frequency Division Multiple Access}
\newacronym{fed4fire}{FED4FIRE+}{Federation 4 Future Internet Research and Experimentation Plus}
\newacronym{fir}{FIR}{Finite Impulse Response}
\newacronym{fit}{FIT}{Future \acrlong{iot}}
\newacronym{fpga}{FPGA}{Field Programmable Gate Array}
\newacronym{fr2}{FR2}{Frequency Range 2}
\newacronym{fs}{FS}{Fast Switching}
\newacronym{fscc}{FSCC}{Flow Sharing Congestion Control}
\newacronym{ftp}{FTP}{File Transfer Protocol}
\newacronym{fw}{FW}{Flow Window}
\newacronym{ge}{GE}{Gaussian Elimination}
\newacronym{gnb}{gNB}{Next Generation Node Base}
\newacronym{nextg}{NextG}{Next Generation}
\newacronym{gop}{GOP}{Group of Pictures}
\newacronym{gpr}{GPR}{Gaussian Process Regressor}
\newacronym{gpu}{GPU}{Graphics Processing Unit}
\newacronym{gtp}{GTP}{GPRS Tunneling Protocol}
\newacronym{gtpc}{GTP-C}{GPRS Tunnelling Protocol Control Plane}
\newacronym{gtpu}{GTP-U}{GPRS Tunnelling Protocol User Plane}
\newacronym{gtpv2c}{GTPv2-C}{\gls{gtp} v2 - Control}
\newacronym{gw}{GW}{Gateway}
\newacronym{harq}{HARQ}{Hybrid Automatic Repeat reQuest}
\newacronym{hetnet}{HetNet}{Heterogeneous Network}
\newacronym{hh}{HH}{Hard Handover}
\newacronym{hol}{HOL}{Head-of-Line}
\newacronym{hqf}{HQF}{Highest-quality-first}
\newacronym{hss}{HSS}{Home Subscription Server}
\newacronym{http}{HTTP}{HyperText Transfer Protocol}
\newacronym{ia}{IA}{Initial Access}
\newacronym{iab}{IAB}{Integrated Access and Backhaul}
\newacronym{ic}{IC}{Incident Command}
\newacronym{ietf}{IETF}{Internet Engineering Task Force}
\newacronym{imsi}{IMSI}{International Mobile Subscriber Identity}
\newacronym{imt}{IMT}{International Mobile Telecommunication}
\newacronym{iot}{IoT}{Internet of Things}
\newacronym{ip}{IP}{Internet Protocol}
\newacronym{itu}{ITU}{International Telecommunication Union}
\newacronym{kpi}{KPI}{Key Performance Indicator}
\newacronym{kpm}{KPM}{Key Performance Measurement}
\newacronym{kvm}{KVM}{Kernel-based Virtual Machine}
\newacronym{los}{LOS}{Line-of-Sight}
\newacronym{lsm}{LSM}{Link-to-System Mapping}
\newacronym{lstm}{LSTM}{Long Short Term Memory}
\newacronym{lte}{LTE}{Long Term Evolution}
\newacronym{lxc}{LXC}{Linux Container}
\newacronym{m2m}{M2M}{Machine to Machine}
\newacronym{mac}{MAC}{Medium Access Control}
\newacronym{manet}{MANET}{Mobile Ad Hoc Network}
\newacronym{mano}{MANO}{Management and Orchestration}
\newacronym{mc}{MC}{Multi-Connectivity}
\newacronym{mcc}{MCC}{Mobile Cloud Computing}
\newacronym{mchem}{MCHEM}{Massive Channel Emulator}
\newacronym{mcs}{MCS}{Modulation and Coding Scheme}
\newacronym{mec}{MEC}{Multi-access Edge Computing}
\newacronym{mec2}{MEC}{Mobile Edge Cloud}
\newacronym{mfc}{MFC}{Mobile Fog Computing}
\newacronym{mgen}{MGEN}{Multi-Generator}
\newacronym{mi}{MI}{Mutual Information}
\newacronym{mib}{MIB}{Master Information Block}
\newacronym{miesm}{MIESM}{Mutual Information Based Effective SINR}
\newacronym{mimo}{MIMO}{Multiple Input, Multiple Output}
\newacronym{ml}{ML}{Machine Learning}
\newacronym{mlr}{MLR}{Maximum-local-rate}
\newacronym[plural=\gls{mme}s,firstplural=Mobility Management Entities (MMEs)]{mme}{MME}{Mobility Management Entity}
\newacronym{mmtc}{mMTC}{Massive Machine-Type Communications}
\newacronym{mmwave}{mmWave}{millimeter wave}
\newacronym{mpdccp}{MP-DCCP}{Multipath Datagram Congestion Control Protocol}
\newacronym{mptcp}{MPTCP}{Multipath TCP}
\newacronym{mr}{MR}{Maximum Rate}
\newacronym{mrdc}{MR-DC}{Multi \gls{rat} \gls{dc}}
\newacronym{mse}{MSE}{Mean Square Error}
\newacronym{mss}{MSS}{Maximum Segment Size}
\newacronym{mt}{MT}{Mobile Termination}
\newacronym{mtd}{MTD}{Machine-Type Device}
\newacronym{mtu}{MTU}{Maximum Transmission Unit}
\newacronym{mumimo}{MU-MIMO}{Multi-user \gls{mimo}}
\newacronym{mvno}{MVNO}{Mobile Virtual Network Operator}
\newacronym{nalu}{NALU}{Network Abstraction Layer Unit}
\newacronym{nas}{NAS}{Non-Access Stratum}
\newacronym{nbiot}{NB-IoT}{Narrow Band IoT}
\newacronym{nfv}{NFV}{Network Function Virtualization}
\newacronym{nfvi}{NFVI}{Network Function Virtualization Infrastructure}
\newacronym{nic}{NIC}{Network Interface Card}
\newacronym{nlos}{NLOS}{Non-Line-of-Sight}
\newacronym{now}{NOW}{Non Overlapping Window}
\newacronym{nsm}{NSM}{Network Service Mesh}
\newacronym[type=hidden]{nr}{NR}{New Radio}
\newacronym{nrf}{NRF}{Network Repository Function}
\newacronym{nsa}{NSA}{Non Stand Alone}
\newacronym{nse}{NSE}{Network Slicing Engine}
\newacronym{nssf}{NSSF}{Network Slice Selection Function}
\newacronym{o2i}{O2I}{Outdoor to Indoor}
\newacronym{oai}{OAI}{OpenAirInterface}
\newacronym{oaicn}{OAI-CN}{\gls{oai} \acrlong{cn}}
\newacronym{oairan}{OAI-RAN}{\acrlong{oai} \acrlong{ran}}
\newacronym{oam}{OAM}{Operations, Administration and Maintenance}
\newacronym{ofdm}{OFDM}{Orthogonal Frequency Division Multiplexing}
\newacronym{olia}{OLIA}{Opportunistic Linked Increase Algorithm}
\newacronym{omec}{OMEC}{Open Mobile Evolved Core}
\newacronym{onap}{ONAP}{Open Network Automation Platform}
\newacronym{onf}{ONF}{Open Networking Foundation}
\newacronym{onos}{ONOS}{Open Networking Operating System}
\newacronym{oom}{OOM}{\gls{onap} Operations Manager}
\newacronym{opnfv}{OPNFV}{Open Platform for \gls{nfv}}
\newacronym{orbit}{ORBIT}{Open-Access Research Testbed for Next-Generation Wireless Networks}
\newacronym{os}{OS}{Operating System}
\newacronym{oss}{OSS}{Operations Support System}
\newacronym{pa}{PA}{Position-aware}
\newacronym{pase}{PASE}{Prioritization, Arbitration, and Self-adjusting Endpoints}
\newacronym{pawr}{PAWR}{Platforms for Advanced Wireless Research}
\newacronym{pbch}{PBCH}{Physical Broadcast Channel}
\newacronym{pcef}{PCEF}{Policy and Charging Enforcement Function}
\newacronym{pcfich}{PCFICH}{Physical Control Format Indicator Channel}
\newacronym{pcrf}{PCRF}{Policy and Charging Rules Function}
\newacronym{pdcch}{PDCCH}{Physical Downlink Control Channel}
\newacronym{pdcp}{PDCP}{Packet Data Convergence Protocol}
\newacronym{pdsch}{PDSCH}{Physical Downlink Shared Channel}
\newacronym{pdf}{PDF}{Probability Density Function}
\newacronym{pf}{PF}{Proportionally Fair}
\newacronym{pgw}{PGW}{Packet Gateway}
\newacronym{phich}{PHICH}{Physical Hybrid ARQ Indicator Channel}
\newacronym{phy}{PHY}{Physical}
\newacronym{pmch}{PMCH}{Physical Multicast Channel}
\newacronym{pmi}{PMI}{Precoding Matrix Indicators}
\newacronym{powder}{POWDER}{Platform for Open Wireless Data-driven Experimental Research}
\newacronym{ppo}{PPO}{Proximal Policy Optimization}
\newacronym{ppp}{PPP}{Poisson Point Process}
\newacronym{prach}{PRACH}{Physical Random Access Channel}
\newacronym{prb}{PRB}{Physical Resource Block}
\newacronym{rbg}{RBG}{Resource Block Group}
\newacronym{rbgs}{RBGs}{Resource Block Groups}
\newacronym{psnr}{PSNR}{Peak Signal to Noise Ratio}
\newacronym{pss}{PSS}{Primary Synchronization Signal}
\newacronym{pucch}{PUCCH}{Physical Uplink Control Channel}
\newacronym{pusch}{PUSCH}{Physical Uplink Shared Channel}
\newacronym{qam}{QAM}{Quadrature Amplitude Modulation}
\newacronym{qci}{QCI}{\gls{qos} Class Identifier}
\newacronym{qoe}{QoE}{Quality of Experience}
\newacronym{qos}{QoS}{Quality of Service}
\newacronym{quic}{QUIC}{Quick UDP Internet Connections}
\newacronym{rach}{RACH}{Random Access Channel}
\newacronym{ran}{RAN}{Radio Access Network}
\newacronym[firstplural=end to endcess Technologies (RATs)]{rat}{RAT}{end to endcess Technology}
\newacronym{rcn}{RCN}{Research Coordination Network}
\newacronym{rec}{REC}{Radio Edge Cloud}
\newacronym{ra}{RA}{Resource Allocation}
\newacronym{red}{RED}{Random Early Detection}
\newacronym{renew}{RENEW}{Reconfigurable Eco-system for Next-generation End-to-end Wireless}
\newacronym{rf}{RF}{Radio Frequency}
\newacronym{rfc}{RFC}{Request for Comments}
\newacronym{rfr}{RFR}{Random Forest Regressor}
\newacronym{ric}{RIC}{RAN Intelligent Controller}
\newacronym{rlc}{RLC}{Radio Link Control}
\newacronym{rlf}{RLF}{Radio Link Failure}
\newacronym{rlnc}{RLNC}{Random Linear Network Coding}
\newacronym{rmr}{RMR}{RIC Message Router}
\newacronym{rmse}{RMSE}{Root Mean Squared Error}
\newacronym{rnis}{RNIS}{Radio Network Information Service}
\newacronym{rr}{RR}{Round Robin}
\newacronym{rrc}{RRC}{Radio Resource Control}
\newacronym{rrm}{RRM}{Radio Resource Management}
\newacronym{rru}{RRU}{Remote Radio Unit}
\newacronym{rs}{RS}{Remote Server}
\newacronym{rsrp}{RSRP}{Reference Signal Received Power}
\newacronym{rsrq}{RSRQ}{Reference Signal Received Quality}
\newacronym{rss}{RSS}{Received Signal Strength}
\newacronym{rssi}{RSSI}{Received Signal Strength Indicator}
\newacronym{rtt}{RTT}{Round Trip Time}
\newacronym{rt}{RT}{real-time}
\newacronym{ru}{RU}{Radio Unit}
\newacronym{rw}{RW}{Receive Window}
\newacronym{rx}{RX}{Receiver}
\newacronym{s1ap}{S1AP}{S1 Application Protocol}
\newacronym{sa}{SA}{standalone}
\newacronym{sack}{SACK}{Selective Acknowledgment}
\newacronym{sap}{SAP}{Service Access Point}
\newacronym{sc2}{SC2}{Spectrum Collaboration Challenge}
\newacronym{scef}{SCEF}{Service Capability Exposure Function}
\newacronym{sch}{SCH}{Secondary Cell Handover}
\newacronym{scoot}{SCOOT}{Split Cycle Offset Optimization Technique}
\newacronym{sctp}{SCTP}{Stream Control Transmission Protocol}
\newacronym{sdap}{SDAP}{Service Data Adaptation Protocol}
\newacronym{sdk}{SDK}{Software Development Kit}
\newacronym{sdm}{SDM}{Space Division Multiplexing}
\newacronym{sdma}{SDMA}{Spatial Division Multiple Access}
\newacronym{sdn}{SDN}{Software-defined Networking}
\newacronym{sdr}{SDR}{Software-defined Radio}
\newacronym{seba}{SEBA}{SDN-Enabled Broadband Access}
\newacronym{sgsn}{SGSN}{Serving GPRS Support Node}
\newacronym{sgw}{SGW}{Service Gateway}
\newacronym{si}{SI}{Study Item}
\newacronym{sib}{SIB}{Secondary Information Block}
\newacronym{sinr}{SINR}{Signal to Interference plus Noise Ratio}
\newacronym{sip}{SIP}{Session Initiation Protocol}
\newacronym{siso}{SISO}{Single Input, Single Output}
\newacronym{sla}{SLA}{Service Level Agreement}
\newacronym{sm}{SM}{Service Model}
\newacronym{smf}{SMF}{Session Management Function}
\newacronym{smo}{SMO}{Service Management and Orchestration}
\newacronym{sms}{SMS}{Short Message Service}
\newacronym{smsgmsc}{SMS-GMSC}{\gls{sms}-Gateway}
\newacronym{snr}{SNR}{Signal-to-Noise-Ratio}
\newacronym{son}{SON}{Self-Organizing Network}
\newacronym{sptcp}{SPTCP}{Single Path TCP}
\newacronym{srb}{SRB}{Service Radio Bearer}
\newacronym{srn}{SRN}{Standard Radio Node}
\newacronym{srs}{SRS}{Sounding Reference Signal}
\newacronym{ss}{SS}{Synchronization Signal}
\newacronym{sss}{SSS}{Secondary Synchronization Signal}
\newacronym{st}{ST}{Spanning Tree}
\newacronym{svc}{SVC}{Scalable Video Coding}
\newacronym{tb}{TB}{Transport Block}
\newacronym{tcp}{TCP}{Transmission Control Protocol}
\newacronym{tdd}{TDD}{Time Division Duplexing}
\newacronym{tdm}{TDM}{Time Division Multiplexing}
\newacronym{tdma}{TDMA}{Time Division Multiple Access}
\newacronym{tfl}{TfL}{Transport for London}
\newacronym{tfrc}{TFRC}{TCP-Friendly Rate Control}
\newacronym{tft}{TFT}{Traffic Flow Template}
\newacronym{tgen}{TGEN}{Traffic Generator}
\newacronym{tip}{TIP}{Telecom Infra Project}
\newacronym{to}{TO}{Telco Operator}
\newacronym{tr}{TR}{Technical Report}
\newacronym{trp}{TRP}{Transmitter Receiver Pair}
\newacronym{ts}{TS}{Technical Specification}
\newacronym{tti}{TTI}{Transmission Time Interval}
\newacronym{ttt}{TTT}{Time-to-Trigger}
\newacronym{tx}{TX}{Transmitter}
\newacronym{uas}{UAS}{Unmanned Aerial System}
\newacronym{uav}{UAV}{Unmanned Aerial Vehicle}
\newacronym{udm}{UDM}{Unified Data Management}
\newacronym{udp}{UDP}{User Datagram Protocol}
\newacronym{udr}{UDR}{Unified Data Repository}
\newacronym{ue}{UE}{User Equipment}
\newacronym{uhd}{UHD}{\gls{usrp} Hardware Driver}
\newacronym{ul}{UL}{Uplink}
\newacronym{um}{UM}{Unacknowledged Mode}
\newacronym{uml}{UML}{Unified Modeling Language}
\newacronym{upa}{UPA}{Uniform Planar Array}
\newacronym{upf}{UPF}{User Plane Function}
\newacronym{urllc}{URLLC}{Ultra Reliable and Low Latency Communications}
\newacronym{usa}{U.S.}{United States}
\newacronym{usim}{USIM}{Universal Subscriber Identity Module}
\newacronym{usrp}{USRP}{Universal Software Radio Peripheral}
\newacronym{utc}{UTC}{Urban Traffic Control}
\newacronym{vim}{VIM}{Virtualization Infrastructure Manager}
\newacronym{vm}{VM}{Virtual Machine}
\newacronym{vnf}{VNF}{Virtual Network Function}
\newacronym{volte}{VoLTE}{Voice over \gls{lte}}
\newacronym{voltha}{VOLTHA}{Virtual OLT HArdware Abstraction}
\newacronym{vr}{VR}{Virtual Reality}
\newacronym{vran}{vRAN}{Virtualized \gls{ran}}
\newacronym{vss}{VSS}{Video Streaming Server}
\newacronym{wbf}{WBF}{Wired Bias Function}
\newacronym{wf}{WF}{Waterfilling}
\newacronym{wlan}{WLAN}{Wireless Local Area Network}
\newacronym{osm}{OSM}{Open Source \gls{nfv} Management and Orchestration}
\newacronym{pnf}{PNF}{Physical Network Function}
\newacronym{drl}{DRL}{Deep Reinforcement Learning}
\newacronym{rl}{RL}{Reinforcement Learning}
\newacronym{mtc}{MTC}{Machine-type Communications}
\newacronym{osc}{OSC}{O-RAN Software Community}
\newacronym{rc}{RC}{RAN Control}
\newacronym{v2x}{V2X}{Vehicle-to-everything}
\newacronym{gbsm}{GBSM}{Geometry-Based Stochastic Model}
\newacronym{gbs}{GBSM}{Geometry-Based Stochastic}
\newacronym{quadriga}{QuaDRiGa}{QUAsi Deterministic RadIo channel GenerAtor}
\newacronym{relu}{ReLU}{Rectified Linear Unit}
\newacronym{mpc}{MPC}{Multipath Component}
\newacronym{nn}{NN}{Neural Network}
\newacronym{sgd}{SGD}{Stochastic Gradient Descent}
\newacronym{cpi}{CPI}{Conservative Policy Iteration}
\newacronym{trpo}{TRPO}{Trust Region Policy Optimization}
\newacronym{mrat}{multi-RAT}{Multi-Radio Access Technology}
\newacronym{confd}{CD}{Conflict Detection}
\newacronym{confr}{CR}{Conflict Resolution}
\newacronym{dcd}{DCD}{Direct Conflict Detection}
\newacronym{icd}{ICD}{Indirect Conflict Detection}
\newacronym{imcd}{ImCD}{Implicit Conflict Detection}
\newacronym{dqn}{DQN}{Deep Q-Network}
\newacronym{smadrl}{SMADRL}{sequential multi-agent deep reinforcement learning}
\newacronym{cmadrl}{CMADRL}{concurrent multi-agent deep reinforcement learning}
\newacronym{tmadrl}{TMADRL}{team multi-agent deep reinforcement learning}
\newacronym{ks}{K-S}{Kolmogorov-Smirnov}
\newacronym{int}{INT}{Integral Area}
\newacronym{pr}{PF}{Proportional Fair}
\newacronym{tm}{TM}{Throughput Maximization}
\newacronym{es}{ES}{Energy Saving}
\newacronym{cov}{COV}{Coefficient of Variation}
\newacronym{rmssd}{RMSSD}{Root Mean Square of Successive Differences}
\newacronym{stdev}{SD}{Standard Deviation}

\glsdisablehyper

\newcommand{\name}{PACIFISTA\xspace}
\newcommand{\oran}{O-RAN\xspace}

\newcommand{\coloran}{Col\oran\xspace}
\newcommand{\scope}{SCOPE\xspace}
\newcommand{\ran}{\gls{ran}\xspace}
\newcommand{\ric}{\gls{ric}\xspace}
\newcommand{\rics}{\glspl{ric}\xspace}
\newcommand{\nearrt}{Near-\gls{rt}\xspace}
\newcommand{\nonrt}{Non-\gls{rt}\xspace}
\newcommand{\ai}{\gls{ai}\xspace}
\newcommand{\apps}{\mathcal{A}}
\newcommand{\params}{\mathcal{P}}
\newcommand{\metrics}{\mathcal{K}}
\newcommand{\kpm}{\gls{kpm}\xspace}
\newcommand{\kpms}{\glspl{kpm}\xspace}
\newcommand{\gp}{G^\mathrm{P}}
\newcommand{\gk}{G^\mathrm{K}}
\newcommand{\ops}{\mathcal{C}}
\newcommand{\confp}{\Pi}
\newcommand{\confa}{\Theta}
\newcommand{\tol}{\delta^{\mathrm{TOL}}}

\newif\ifexttikz
\exttikzfalse

\ifexttikz
	\usetikzlibrary{external}
\fi

\newlength\fheight
\newlength\fwidth
\usepackage{collcell}
\usepackage{hhline}
\usepackage{pgf}
\usepackage{multirow}

\def\colorModel{hsb} %
\newcommand\ColCell[1]{
	\pgfmathparse{#1<0.5?1:0}  %
	\pgfmathsetmacro\compA{(1-#1)/3} %
	\pgfmathsetmacro\compB{0.65} %
	\pgfmathsetmacro\compC{0.9} %

	\edef\x{\noexpand\centering\noexpand\cellcolor[\colorModel]{\compA,\compB,\compC}}\x #1
}

\newcolumntype{E}{>{\collectcell\ColCell}m{2.3em}<{\endcollectcell}}
\newcolumntype{F}{>{\collectcell\ColCell}m{1.8em}<{\endcollectcell}}

\usepackage{listings}

\definecolor{codegray}{rgb}{0.25,0.25,0.25}
\definecolor{codepurple}{rgb}{0.58,0,0.82}
\definecolor{pPeriwinkle}{rgb}{204, 204, 255}
\definecolor{mBlue}{rgb}{25, 25, 112}

\lstdefinestyle{mystyle-yaml}{
	basicstyle=\color{black}\ttfamily\scriptsize,
	commentstyle=\color{gray},
	keywordstyle=\color{codePurple},
	numberstyle=\tiny\color{codeGray},
	stringstyle=\color{codePurple},
	rulecolor=\color{black},
	breakatwhitespace=true,
	breaklines=true,
	captionpos=b,
	frame=tb,
	keepspaces=true,
	numbers=left,
	numbersep=5pt,
	showspaces=false,
	showstringspaces=false,
	showtabs=false,
	tabsize=2,
	xleftmargin=10pt,
	belowskip=-10pt,
	float=htbp,  %
}

\lstdefinestyle{mystyle-yaml-leo}{
	commentstyle=\color{gray},
	keywordstyle=\color{purple},
	numberstyle=\tiny\color{codegray},
	stringstyle=\color{codepurple},
	basicstyle=\color{Periwinkle}\ttfamily\scriptsize,
	rulecolor=\color{black},
	breakatwhitespace=true,
	breaklines=true,
	captionpos=b,
	frame=tb,
	keepspaces=true,
	numbers=left,
	numbersep=5pt,
	showspaces=false,
	showstringspaces=false,
	showtabs=false,
	tabsize=2,
	xleftmargin=10pt,
}

\lstdefinelanguage{yaml}{
	alsoletter={-},
	keywords={true,false,null,y,n,-},
	sensitive=false,
	comment=[l]{\#},
	morecomment=[s]{/*}{*/},
	moredelim=[l][\color{orange}]{\&},
	moredelim=[l][\color{magenta}]{*},
	morestring=[b]',
	morestring=[b]",
}

\theoremstyle{plain}
\newtheorem{theorem}{Theorem}

\theoremstyle{definition}
\newtheorem{definition}[theorem]{Definition}
\newtheorem*{definition*}{Definition}

\theoremstyle{remark}
\newtheorem{remark}{Remark}
\newtheorem*{remark*}{Remark}

\AtBeginDocument{%
\providecommand\BibTeX{{%
\normalfont B\kern-0.5em{\scshape i\kern-0.25em b}\kern-0.8em\TeX}}}

\usepackage{pifont}%

\begin{document}

\title{\name: Conflict Evaluation and Management in Open RAN}

\author{\IEEEauthorblockN{
		Pietro Brach del Prever,~\IEEEmembership{Student Member, IEEE},
		Salvatore D'Oro,~\IEEEmembership{Member, IEEE},\\
		Leonardo Bonati,~\IEEEmembership{Member, IEEE},
		Michele Polese,~\IEEEmembership{Member, IEEE},
		Maria~Tsampazi,~\IEEEmembership{Student~Member,~IEEE},
		Heiko Lehmann,
		Tommaso Melodia,~\IEEEmembership{Fellow, IEEE}}

	\thanks{P. Brach del Prever, S. D'Oro, L. Bonati, M. Polese, M. Tsampazi, and T. Melodia are with the Institute for the Wireless Internet of Things, Northeastern University, Boston, MA, U.S.A. E-mail: \{brachdelprever.p, s.doro, l.bonati, m.polese, tsampazi.m, melodia\}@northeastern.edu.
		Heiko Lehman is with Deutsche Telekom AG, T-Labs, 10781 Berlin, Germany. Email: h-lehmann@telekom.de.}
	\thanks{This work was partially supported by Deutsche Telekom, by the U.S. National Science Foundation under grant CNS-1925601, and by OUSD(R\&E) through Army Research Laboratory Cooperative Agreement Number W911NF-24-2-0065. The views and conclusions contained in this document are those of the authors and should not be interpreted as representing the official policies, either expressed or implied, of the Army Research Laboratory or the U.S. Government. The U.S. Government is authorized to reproduce and distribute reprints for Government purposes notwithstanding any copyright notation herein.}
}

\maketitle

\begin{abstract}
	The \oran ALLIANCE is defining architectures, interfaces, operations, and security requirements for cellular networks based on Open \ran principles. In this context,  \oran introduced the \rics to enable dynamic control of cellular networks via data-driven applications referred to as rApps and xApps. \rics enable for the first time truly intelligent and self-organizing cellular networks. However, enabling the execution of many \ai algorithms making autonomous control decisions to fulfill diverse (and possibly conflicting) goals poses unprecedented challenges. For instance, the execution of one xApp aiming at maximizing throughput and one aiming at minimizing energy consumption would inevitably result in diametrically opposed resource allocation strategies. Therefore,  conflict management becomes a crucial component of any functional intelligent \oran system.
	This article studies the problem of conflict mitigation in \oran and proposes \name, a framework to detect, characterize, and mitigate conflicts generated by \oran applications that control \ran parameters.
	\name leverages a profiling pipeline to tests \oran applications in a sandbox environment,
	and combines hierarchical graphs with statistical models to detect the existence of conflicts and evaluate their severity. Experiments on Colosseum and OpenRAN Gym demonstrate \name's ability to predict conflicts and provide valuable information before potentially conflicting xApps are deployed in production systems.
	We use \name to demonstrate that users can experience a 16\% throughput loss even in the case of xApps with similar goals, and that applications with conflicting goals might cause severe instability and result in up to 30\% performance degradation.
	We also show that \name~can help operators to identify conflicting applications and maintain performance degradation below a tolerable threshold.
\end{abstract}

\begin{IEEEkeywords}
	Conflict Management, O-RAN, Open RAN, 5G, 6G.
\end{IEEEkeywords}

\begin{picture}(0,0)(10,-570)
	\put(0,0){
		\put(0,0){\footnotesize This paper has been accepted for publication on IEEE Transactions on Mobile Computing.}
		\put(0,-10){
			\scriptsize \textcopyright~2025 IEEE. Personal use of this material is permitted. Permission from IEEE must be obtained for all other uses, in any current or future media, including}
		\put(0, -17){
			\scriptsize reprinting/republishing this material for advertising or promotional purposes, creating new collective works, for resale or redistribution to servers or lists,}
		\put(0, -24){
			\scriptsize or reuse of any copyrighted component of this work in other works.}
	}
\end{picture}

\glsresetall

\vspace{-0.2cm}
\section{Introduction}\label{sec:introduction}

The Open \ran paradigm is spearheading the revolution in the telco ecosystem by promoting open, programmable, virtualized, multi-vendor, disaggregated cellular architectures. In this context, the \oran ALLIANCE---a consortium of vendors, operators, integrators, and academic partners---is specifying Open \ran architectures, interfaces, operations, and security requirements necessary to realize the Open \ran vision~\cite{polese2023understanding}.

One of the most disrupting technologies in \oran are the \rics, i.e., the \nonrt \ric and the \nearrt \ric. Both host intelligent applications that execute inference tasks (e.g., monitoring, control, forecasting). The former is designed to operate on timescales above $1$\:s via so-called \textit{rApps}, while the latter hosts \textit{xApps} that perform tasks on timescales between $10$\:ms and $1$\:s~\cite{oran-wg1-use-cases}.
This enables dynamic and efficient policy control to reconfigure the \ran and achieve bespoke operator goals while adapting to varying demand and load.

\begin{figure}[t!]
	\setlength\abovecaptionskip{0pt}
	\setlength\belowcaptionskip{0pt}
	\centering
	\setlength\fwidth{\columnwidth}
	\setlength\fheight{.7\columnwidth}
	\begin{tikzpicture}

	\begin{axis}[
			width=0.52\fwidth,
			height=0.35\fheight,
			at={(0\linewidth,0\linewidth)},
			scale only axis,
			xmin=0,
			xmax=270,
			xtick={0,45,...,270},
			xlabel style={font=\scriptsize\color{white!15!black}},
			xlabel={Time [s]},
			ymin=0,
			ymax=40,
			ylabel style={font=\scriptsize\color{white!15!black}},
			ylabel={PRBs},
			axis background/.style={fill=white},
			axis x line*=bottom,
			axis y line*=left,
			xmajorgrids,
			ymajorgrids,
			ylabel shift=-4pt,
			xlabel shift=-3pt,
			enlargelimits=false
		]
		\addplot [color=Paired-B, forget plot]
		table[row sep=crcr]{
				0.25	30\\
				0.5	30\\
				0.75	24\\
				2.5	24\\
				2.75	27\\
				4.5	27\\
				4.75	30\\
				6.5	30\\
				6.75	27\\
				8.5	27\\
				8.75	33\\
				10.5	33\\
				10.75	27\\
				16.5	27\\
				16.75	33\\
				18.5	33\\
				18.75	30\\
				20.5	30\\
				20.75	27\\
				22.5	27\\
				22.75	30\\
				24.5	30\\
				24.75	27\\
				30.5	27\\
				30.75	30\\
				32.5	30\\
				32.75	27\\
				34.5	27\\
				34.75	30\\
				40.5	30\\
				40.75	33\\
				42.5	33\\
				42.75	27\\
				48.5	27\\
				48.75	30\\
				52.5	30\\
				52.75	27\\
				54.5	27\\
				54.75	30\\
				60.5	30\\
				60.75	27\\
				64.5	27\\
				64.75	30\\
				66.5	30\\
				66.75	27\\
				68.5	27\\
				68.75	30\\
				74.5	30\\
				74.75	27\\
				76.5	27\\
				76.75	30\\
				80.5	30\\
				80.75	27\\
				85	27\\
				85.25	30\\
				87	30\\
				87.25	27\\
				90	27\\
				90.25	3\\
				93	3\\
				93.25	6\\
				97	6\\
				97.25	3\\
				99	3\\
				99.25	6\\
				101	6\\
				101.2	3\\
				105	3\\
				105.2	6\\
				113	6\\
				113.2	3\\
				117.5	3\\
				117.8	6\\
				127.5	6\\
				127.8	3\\
				135.5	3\\
				135.8	6\\
				139.5	6\\
				139.8	3\\
				145.5	3\\
				145.8	6\\
				147.5	6\\
				147.8	3\\
				149.5	3\\
				149.8	6\\
				151.5	6\\
				151.8	3\\
				155.5	3\\
				155.8	6\\
				157.5	6\\
				157.8	3\\
				159.5	3\\
				159.8	9\\
				161.5	9\\
				161.8	6\\
				165.5	6\\
				165.8	3\\
				173.5	3\\
				173.8	6\\
				177.5	6\\
				177.8	3\\
				180	3\\
				180.2	6\\
				181	6\\
				181.2	30\\
				182	30\\
				182.2	3\\
				183	3\\
				183.2	33\\
				184.5	33\\
				184.8	3\\
				185	3\\
				185.2	27\\
				186.5	27\\
				186.8	6\\
				187	6\\
				187.2	27\\
				188.5	27\\
				188.8	6\\
				189	6\\
				189.2	30\\
				190.5	30\\
				190.8	3\\
				191	3\\
				191.2	24\\
				192.5	24\\
				192.8	3\\
				193	3\\
				193.2	27\\
				194.5	27\\
				194.8	3\\
				195	3\\
				195.2	27\\
				196.5	27\\
				196.8	6\\
				197	6\\
				197.2	27\\
				198.5	27\\
				198.8	6\\
				199	6\\
				199.2	30\\
				200.5	30\\
				200.8	6\\
				201	6\\
				201.2	30\\
				202.5	30\\
				202.8	3\\
				203	3\\
				203.2	30\\
				204.5	30\\
				204.8	3\\
				205	3\\
				205.2	27\\
				206.5	27\\
				206.8	3\\
				207	3\\
				207.2	30\\
				208.5	30\\
				208.8	6\\
				209	6\\
				209.2	27\\
				210.5	27\\
				210.8	3\\
				211	3\\
				211.2	30\\
				212.5	30\\
				212.8	3\\
				213	3\\
				213.2	30\\
				214.5	30\\
				214.8	3\\
				215	3\\
				215.2	27\\
				216.5	27\\
				216.8	3\\
				217	3\\
				217.2	27\\
				218.5	27\\
				218.8	3\\
				219	3\\
				219.2	27\\
				220.5	27\\
				220.8	3\\
				221	3\\
				221.2	27\\
				222.5	27\\
				222.8	3\\
				223	3\\
				223.2	27\\
				224.5	27\\
				224.8	6\\
				225	6\\
				225.2	30\\
				226.5	30\\
				226.8	3\\
				227	3\\
				227.2	27\\
				228.5	27\\
				228.8	3\\
				229	3\\
				229.2	30\\
				230.5	30\\
				230.8	6\\
				231	6\\
				231.2	30\\
				232.5	30\\
				232.8	6\\
				233	6\\
				233.2	30\\
				234.5	30\\
				234.8	3\\
				235	3\\
				235.2	30\\
				236.5	30\\
				236.8	3\\
				237	3\\
				237.2	27\\
				238.5	27\\
				238.8	3\\
				239	3\\
				239.2	30\\
				240.5	30\\
				240.8	6\\
				241	6\\
				241.2	30\\
				242.5	30\\
				242.8	6\\
				243	6\\
				243.2	27\\
				244.5	27\\
				244.8	3\\
				245	3\\
				245.2	30\\
				246.5	30\\
				246.8	3\\
				247	3\\
				247.2	27\\
				248.5	27\\
				248.8	6\\
				249	6\\
				249.2	30\\
				250.5	30\\
				250.8	3\\
				251	3\\
				251.2	27\\
				252.5	27\\
				252.8	3\\
				253	3\\
				253.2	27\\
				254.5	27\\
				254.8	6\\
				255	6\\
				255.2	27\\
				256.5	27\\
				256.8	3\\
				257	3\\
				257.2	27\\
				258.5	27\\
				258.8	3\\
				259	3\\
				259.2	27\\
				260.5	27\\
				260.8	6\\
				261	6\\
				261.2	30\\
				262.5	30\\
				262.8	6\\
				263	6\\
				263.2	27\\
				264.5	27\\
				264.8	6\\
				265	6\\
				265.2	27\\
				266.5	27\\
				266.8	6\\
				267	6\\
				267.2	30\\
				268.5	30\\
				268.8	3\\
				269	3\\
				269.2	27\\
				270	27\\
			};

		\draw[anchor=center,{stealth[]}-{stealth[]}] (0, 37) -- (45, 37) node[font=\tiny,fill=white,draw=black,inner sep=1.2pt, anchor=center] {TM} -- (90, 37);

		\addplot [color=black, forget plot]
		table[y index=1, row sep=crcr]{
				90	0\\
				90	50\\
			};

		\draw[anchor=center,{stealth[]}-{stealth[]}] (90, 37) -- (135, 37) node[font=\tiny,fill=white,draw=black,inner sep=1.2pt, anchor=center] {ES} -- (180, 37);

		\addplot [color=black, forget plot]
		table[y index=1, row sep=crcr]{
				180	0\\
				180	50\\
			};

		\draw[anchor=center,{stealth[]}-{stealth[]}] (180, 37) -- (225, 37) node[font=\tiny,fill=white,draw=black,inner sep=1.2pt, anchor=center] {TM+ES} -- (270, 37);

		\addplot [color=black, forget plot]
		table[y index=1, row sep=crcr]{
				270	0\\
				270	50\\
			};

	\end{axis}

	\begin{axis}[
			width=0.25\fwidth,
			height=0.35\fheight,
			at={(0.62\fwidth,0\fheight)},
			scale only axis,
			xmin=0,
			xmax=4,
			xtick={1, 2, 3},
			xticklabels={{TM},{ES},{TM+ES}},
			x tick label style={
					rotate=90, anchor=east,
					tick style={font=\tiny, draw=none}},
			ymin=0,
			ymax=6,
			ylabel style={font=\scriptsize\color{white!15!black}},
			ylabel={Throughput [Mbps]},
			axis background/.style={fill=white},
			xmajorgrids=false,
			ymajorgrids,
			ylabel shift=-5pt,
			xlabel shift=-3pt,
			enlargelimits=false
		]

		\addplot [color=black, forget plot,
		]
		table[row sep=crcr]{%
				1	4.229\\
				1	5.866\\
			};
		\addplot [color=black, forget plot,
		]
		table[row sep=crcr]{%
				2	0.705\\
				2	1.273\\
			};
		\addplot [color=black, forget plot,
		]
		table[row sep=crcr]{%
				3	3.637\\
				3	5.962\\
			};
		\addplot [color=black, forget plot,
		]
		table[row sep=crcr]{%
				1	0.3058\\
				1	2.57\\
			};
		\addplot [color=black, forget plot,
		]
		table[row sep=crcr]{%
				2	0.04873\\
				2	0.3114\\
			};
		\addplot [color=black, forget plot,
		]
		table[row sep=crcr]{%
				3	0.2044\\
				3	1.644\\
			};
		\addplot [color=black, forget plot]
		table[row sep=crcr]{%
				0.8875	5.866\\
				1.112	5.866\\
			};
		\addplot [color=black, forget plot]
		table[row sep=crcr]{%
				1.888	1.273\\
				2.112	1.273\\
			};
		\addplot [color=black, forget plot]
		table[row sep=crcr]{%
				2.888	5.962\\
				3.112	5.962\\
			};
		\addplot [color=black, forget plot]
		table[row sep=crcr]{%
				0.8875	0.3058\\
				1.113	0.3058\\
			};
		\addplot [color=black, forget plot]
		table[row sep=crcr]{%
				1.888	0.04873\\
				2.112	0.04873\\
			};
		\addplot [color=black, forget plot]
		table[row sep=crcr]{%
				2.888	0.2044\\
				3.112	0.2044\\
			};
		\addplot [color=Paired-B, thick, forget plot]
		table[row sep=crcr]{%
				0.775	2.57\\
				0.775	4.229\\
				1.225	4.229\\
				1.225	2.57\\
				0.775	2.57\\
			};
		\addplot [color=Paired-B, thick, forget plot]
		table[row sep=crcr]{%
				1.775	0.3114\\
				1.775	0.705\\
				2.225	0.705\\
				2.225	0.3114\\
				1.775	0.3114\\
			};
		\addplot [color=Paired-B, thick, forget plot]
		table[row sep=crcr]{%
				2.775	1.644\\
				2.775	3.637\\
				3.225	3.637\\
				3.225	1.644\\
				2.775	1.644\\
			};
		\addplot [color=Paired-F, thick, forget plot]
		table[row sep=crcr]{%
				0.775	3.907\\
				1.225	3.907\\
			};
		\addplot [color=Paired-F, thick, forget plot]
		table[row sep=crcr]{%
				1.775	0.4678\\
				2.225	0.4678\\
			};
		\addplot [color=Paired-F, thick, forget plot]
		table[row sep=crcr]{%
				2.775	2.476\\
				3.225	2.476\\
			};
		\addplot [color=Paired-F, thick, forget plot]
		table[row sep=crcr]{%
				90	0\\
				90	10\\
			};
		\addplot [color=Paired-F, thick, forget plot]
		table[row sep=crcr]{%
				180	0\\
				180	10\\
			};
		\addplot [color=Paired-F, thick, forget plot]
		table[row sep=crcr]{%
				270	0\\
				270	10\\
			};
	\end{axis}
\end{tikzpicture}
	\caption{{Impact of conflicts on performance. The xApps for \gls{tm} and for \gls{es} are first run separately and then together. Left: assigned \glspl{prb} for \gls{embb} slice. Right: Measured throughput statistics.}}
	\label{fig:motivational}
	\vspace{-0.1cm}
\end{figure}
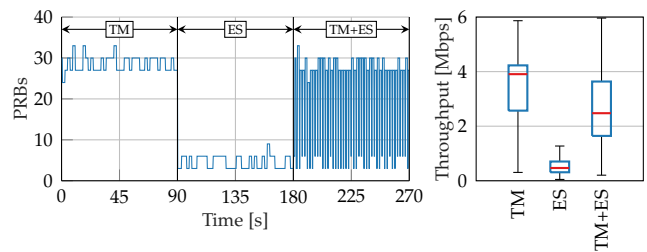

\oran paves the way to self-optimizing cellular networks rooted in data-driven policy customization based on real-time \ran performance.
In this way, \ai control enables benefits such as improved efficiency and performance, and reduced energy consumption, among others~\cite{pamuklu2021energy}.
However, the co-existence of a multitude of \ai-based algorithms taking autonomous decisions to achieve diverse goals (e.g., maximizing performance or minimizing energy consumption) exposes the network to conflicting control policies. An illustrative example of a conflict between two control policies is that of a \gls{tm} xApp---trying to maximize \gls{dl} throughput for a \gls{embb} slice---and an \gls{es} xApp---minimizing energy consumption of the \ran. In Fig.~\ref{fig:motivational} (left), we report the control policies (i.e., \glspl{prb} allocation for the \gls{embb} slice) computed by the two xApps when executing on a real-world \oran testbed. We notice how the two xApps allocate \glspl{prb} differently and according to their individual goals. We also notice that when executing at the same time (i.e., the \gls{tm} + ES case), the conflicting goals generate unstable control policies that cause an oscillatory behavior. These oscillating control policies are undesirable, because both apps try to achieve their goals---i.e., maximize throughput by giving a lot of resources, or maximize energy efficiency, by assigning very few resources---but none manages to maintain the network configurations that they want to set. Consequently, this behavior hinders performance. Specifically, Fig.~\ref{fig:motivational} (right) shows how xApp \gls{tm} delivers $\approx4$~Mbps median throughput values, while xApp ES saves energy by maintaining the number of allocated \glspl{prb} low ($\approx$5 \glspl{prb} out of 50). When both \gls{tm} and ES xApps coexist at the same time, throughput drops by $\approx$50\% compared to xApp \gls{tm}, while resource utilization increases compared to xApp ES, which is a behavior that goes against the intents of both xApps.

\ifexttikz
	\tikzsetnextfilename{motivational_figure}
\fi

The problem of conflict management in the Open \ran~is quite broad. The \oran~ALLIANCE recognizes several conflicts that include cell ON/OFF, beamforming, handover, antenna tilt, traffic steering, and many others and mentions the importance of conflict management frameworks such as \name~in \cite{oran-wg3-con-mit}. Indeed, it is not tolerable that an xApp triggers a handoff of certain users to another cell, and the target cell gets turned off by another xApp trying to save energy.
Industry is clearly showing that conflicts are a great deal in \oran and require proper solutions to mitigate performance degradation and prevent conflicts that might cause service disruption, outages, and even monetary loss. These frameworks indeed increase the level of complexity of the network, but are necessary to ensure its continuous and reliable operations.

In principle, one could avoid the need for conflict management frameworks via conflict avoidance~\cite{d2022orchestran}. For instance, one could decide not to deploy xApp ES when xApp \gls{tm} is active. However, granted this approach might be feasible for small \ran deployments, it would limit the benefits of the \ric, as some conflicting applications might be able to coexist under certain operational conditions. For example, xApp \gls{tm} would increase resource utilization when there is user demand, but would save resources when there is no demand, which aligns with the goal of xApp ES.

Another naive approach would consider a centralized entity overseeing the entire network. Unfortunately, this approach is impractical as it needs a unified algorithm to control thousands of \ran components and functionalities simultaneously and in real-time, which is unfeasible due to the combinatorial number of actions and network states to be explored. On the other hand, a distributed intelligence approach utilizing multiple xApps and rApps, each controlling specific parts of the network to achieve individual goals, provides a more practical, scalable, and programmable solution.
However, how to guarantee that this fabric of \ai-based, multi-vendor xApps and rApps makes decisions without generating conflicting control policies---that might result in performance degradation---is still unclear, especially since
conflicts can be diverse, observable only at certain timescales, or affecting different network components and \kpms.

Given the complexity and significance of the issue, conflict management has emerged as a key area of interest within the community, offering a crucial tool for enabling and promoting the adoption of \oran.
Specifically, the \oran ALLIANCE has classified conflicts into direct, indirect and implicit ones (discussed in detail later in the paper).
Preliminary efforts in \oran conflict management include detecting conflicting control policies in real time~\cite{adamczyk2023challenges,adamczyk2023conflict}, performing on-line ``deconfliction''~\cite{wadud2024qacm}, coordinating \ai-based xApps and rApps via team learning to reduce the occurrence of conflicts~\cite{iturria2022multi,zhang2022team}, as well as orchestrating xApp and rApp selection and deployment to avoid conflicts~\cite{d2022orchestran}.
While these works illustrate concrete efforts to develop solutions that can address and mitigate conflicts in \oran, they do not fully examine the nuances of conflict severity and their impact on \glspl{kpm}. Additionally, they do not consider the possibility that certain conflicts may only become apparent under specific operational conditions (e.g., xApps \gls{tm} and \gls{es} may exhibit a similar behavior when network load is low).

\begin{figure}[t!]
	\centering
	\includegraphics[width=\linewidth]{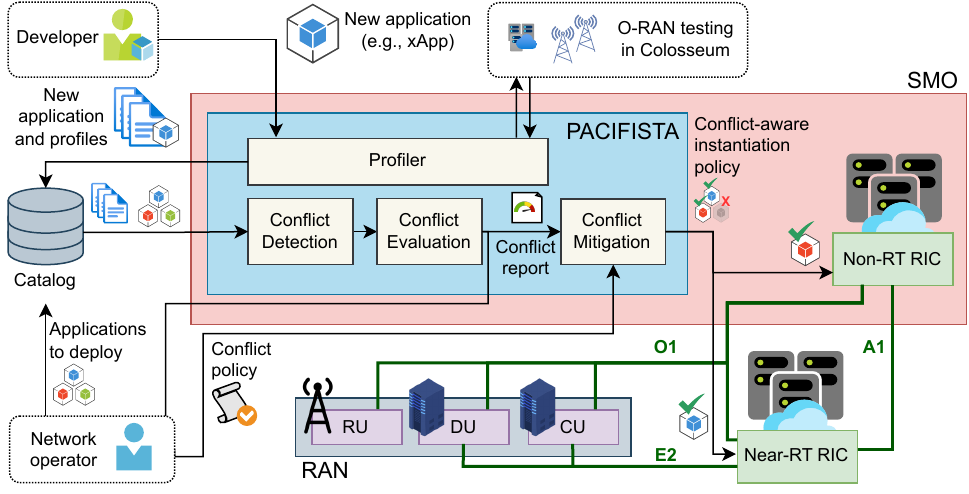}
	\vspace{-0.5cm}
	\caption{\name architecture and workflow.}
	\label{fig:system_architecture}
	\vspace{-0.1cm}
\end{figure}

\textbf{Main contributions.}~In this paper, we fill this gap by proposing \name, a framework to characterize, and evaluate direct, indirect and implicit conflicts in \oran. Specifically:

\begin{itemize}[noitemsep,topsep=0pt]
	\item We design and present \name, an empirical and formal framework that draws on statistical information from xApps, rApps and dApps~\cite{d2022dapps} to detect, characterize and mitigate conflicts that might arise between different applications. The architecture of \name is illustrated in Fig.~\ref{fig:system_architecture}. \name combines the use of a hierarchical graph with the knowledge of the statistical behavior of \oran applications to identify relationships between control parameters and \kpms. This enables \name to: (i)~determine whether two or more applications will generate conflicts; and (ii)~quantify the severity of such conflicts.
	\item We design a profiling pipeline to test \oran applications across a set of predefined sandbox tests in different operational conditions. The pipeline generates a mathematical model to infer the existence and severity of conflicts by leveraging the statistical data generated during the profiling phase. We present several statistical indicators and tools that can be used to assess the severity of each conflict and identify affected \kpms and control parameters.
	\item We demonstrate the capabilities of \name via experiments on the Colosseum wireless network emulator and the OpenRAN Gym~\cite{bonati2023openran} platform with real xApps, showing that \name can predict the occurrence of conflicts and provide accurate information on which \kpms will be affected and to what extent. In this manner, the a-priori knowledge offered by \name can be used to evaluate conflicts prior to the deployment of \oran applications, thereby facilitating  informed deployment decisions regarding which xApps/rApps to deploy to mitigate conflicts.
\end{itemize}

\vspace{-0.2cm}
\section{Related Work} \label{sec:related}

Conflict mitigation is a problem that applies in general to distributed systems with multiple agents having access and control over a set of objects/systems~\cite{lupu1999conflicts}. It is generally tackled using a combination of game theory, access control and coordination mechanisms~\cite{wadud2024qacm,tessier2005conflicting,cuppens2007high,lee2010conflict}. In the O-RAN context, the control surface is represented by the wireless portion of the network (e.g., resource management, user mobility, node scaling, among others), whose dynamics are stochastic in nature and hard to predict. This makes conflict management and resolution in \oran a substantially different problem than those involved in supply chain, control, and information systems where dynamics are instead more predictable and slowly varying, thus making the approaches mentioned above unsuitable and ineffective. Next, we provide a primer on the \oran architecture and discuss related work in conflict mitigation.

\textbf{\oran---A Primer.} The \oran architecture combines a disaggregated \ran with the \rics, deployed on an infrastructure composed of servers, hardware accelerators, and virtualization solutions, collectively referred to as the O-Cloud. Fig.~\ref{fig:system_architecture} provides a logical diagram of this architecture. The disaggregated \ran features \glspl{gnb} split into a \gls{ru}, \gls{du}, and \gls{cu}, implementing different portions of the protocol stack. \glspl{du} and \glspl{cu} are connected to the \nearrt \ric through the E2 interface, while all \glspl{gnb} components connect to the network \gls{smo} framework via the O1 interface. The \gls{smo} embeds the \nonrt \ric, which connects to the \nearrt \ric via the A1 interface. More details on the \oran architecture are discussed in~\cite{polese2023understanding,oran-wg1-arch-spec}.

\textbf{Conflict Management in \oran.}
A systematic analysis of the challenges of conflict control in the \ran and of the strategies proposed to address them is presented in \cite{adamczyk2023challenges}.
With respect to literature on conflicts in wireless systems, which primarily concerns the avoidance of interference in ad hoc networks~\cite{li2017conflict,chen2006modeling}, the focus in~\cite{adamczyk2023challenges} and in our paper is on avoiding conflicting configurations in the \ran protocol stack. Specifically, preventive conflict mitigation activities should reliably detect conflicts, also in a network with dynamic conditions, provide optimal conflict resolution and methodologies for testing, and evaluate conflict mitigation methods.

A conflict detection and mitigation framework for xApps at the \nearrt \ric is presented in~\cite{adamczyk2023conflict}. Differently from \name, this framework only detects conflicts on xApps already deployed on the \oran infrastructure. In light of the framework presented in~\cite{adamczyk2023conflict},~\cite{wadud2024qacm} proposes a Quality-of-Service Aware Conflict Mitigation method that identifies an optimal equilibrium point for all xApps while ensuring satisfaction of Quality of Service requirements. The framework, however, presents two potential areas for improvement, which the authors intend to address in future work. First, the framework is heavily dependent on \kpm prediction, which is a complex task due to the dynamic and complex nature of the network. Secondly, it provides validation exclusively through Python-based simulated experiments based on a simplified model for the network \kpm prediction.

A framework for orchestration of the deployment of \oran applications is presented in~\cite{d2022orchestran}. This framework performs conflict avoidance only by ensuring that conflicting applications are not concurrently deployed. However, this work does not perform conflict detection and mitigation, as \name instead does.
A team learning-based strategy to reduce and eliminate conflicts among xApps in the \nearrt \ric is defined in~\cite{zhang2022team}. In this approach, xApps learn to cooperate and avoid conflicts using a \gls{dqn} architecture. However, this solution requires coordination during training, which might not always be the case in \oran, where multiple vendors publish their individually trained xApps and rApps.
Building on this work, a case study of multi-agent team learning for xApps controlling different \ran parameters is presented in~\cite{iturria2022multi}, where authors present a framework that mitigates conflicts in real-time, but does not reduce conflict occurrence by regulating xApp deployment, as \name~instead does.

In~\cite{yungaicela2024misconfiguration}, the authors provide an overview of how conflicts can generate misconfigurations in \oran, and discuss how conflicts can impact all layers of the protocol stack and how this can result in increased energy consumption, performance degradation and instability. They also present a solution to detect conflicts but limit their study to direct and indirect conflicts, while mentioning that detection of implicit conflicts requires a more complex and data-driven solution, such as the one we present in this paper.

Lastly, \cite{zolghadr2024learning} propose an appraoach based on neural networks to learn and construct graphs that describe conflicts.

Compared to the previous literature on the topic, our goal is to design and develop a full-fledged mathematical and operational framework that embeds pipelines to characterize individual \oran applications, and leverage statistical analysis to accurately evaluate the severity of conflicts and compute mitigation strategies both prior to \oran application execution, and at run time.

\vspace{-0.2cm}
\section{Modeling Conflicts in \oran} \label{sec:conflicts}

\subsection{Definitions of Conflicts}\label{sec:conflictdefinition}

We introduce here the three classes of conflicts identified by the \oran Alliance~\cite{oran-wg3-con-mit}.

\begin{figure}[t!]
	\centering
	\newcommand*{\gwidth}{1}
	\newcommand*{\gheight}{1.5}

	\ifexttikz\tikzsetnextfilename{three_graphs_1}\fi
	\begin{subfigure}[t]{0.32\linewidth}
		\centering
		\scalebox{0.8}{\usepgfplotslibrary{patchplots,groupplots,colorbrewer}
\pgfplotsset{
	compat=newest,
	plot coordinates/math parser=false,
	every tick label/.style={font=\tiny},
	colormap/Paired
}

\begin{tikzpicture}[auto]

	\begin{scope}[nodes={circle,draw,fill opacity=0.7}]
		\node[fill=Paired-A] (a1) at (-\gwidth/2,0) {$a_1$};
		\node[fill=Paired-A] (a2) at (\gwidth/2, 0) {$a_2$};
		\node[fill=Paired-G!80] (p1) at (-\gwidth, -\gheight) {$p_1$};
		\node[fill=Paired-G!80, draw=Paired-F] (p2) at (0, -\gheight) {$p_2$};
		\node[fill=Paired-G!80] (p3) at (\gwidth, -\gheight) {$p_3$};
	\end{scope}

	\begin{scope}[>=stealth']
		\draw[->] (a1) -- (p1);
		\draw[->, draw=Paired-F] (a1) -- (p2);
		\draw[->, draw=Paired-F] (a2) -- (p2);
		\draw[->] (a2) -- (p3);
	\end{scope}

	\begin{scope}[every node/.style={fill=white, font=\tiny}]
		\node[align=center, font=\bf, rotate=90] at (-\gwidth*1.5-0.1,0) {Apps};
		\draw[dashed] (-\gwidth*1.5,-\gheight/2)--(\gwidth*1.3,-\gheight/2);
		\node[align=center, font=\bf, rotate=90] at (-\gwidth*1.5-0.1,-\gheight) {Params};
	\end{scope}

	\begin{scope}[every node/.style={font=\tiny\bf,
					fill=white
				}]
		\node[align=center, text=Paired-F, inner sep=1pt] at (0,-\gheight/2) {DIRECT\\ CONFLICT};
	\end{scope}

\end{tikzpicture}}
		\caption{Direct conflict.}
		\label{fig:direct}
	\end{subfigure}
	\ifexttikz\tikzsetnextfilename{three_graphs_2}\fi
	\begin{subfigure}[t]{0.32\linewidth}
		\centering
		\scalebox{0.8}{\usepgfplotslibrary{patchplots,groupplots,colorbrewer}
\pgfplotsset{
	compat=newest,
	plot coordinates/math parser=false,
	every tick label/.style={font=\tiny},
	colormap/Paired
}

\begin{tikzpicture}[auto]

	\begin{scope}[nodes={circle,draw,fill opacity=0.7}]
		\node[fill=Paired-G!80] (p1) at (-\gwidth/2,0) {$p_1$};
		\node[fill=Paired-G!80] (p2) at (\gwidth/2, 0) {$p_2$};
		\node[fill=Paired-I!80] (k1) at (-\gwidth, -\gheight) {$k_1$};
		\node[fill=Paired-I!80, draw=Paired-F] (k2) at (0, -\gheight) {$k_2$};
		\node[fill=Paired-I!80] (k3) at (\gwidth, -\gheight) {$k_3$};
	\end{scope}

	\begin{scope}[>=stealth']
		\draw[->] (p1) -- (k1);
		\draw[->, draw=Paired-F] (p1) -- (k2);
		\draw[->, draw=Paired-F] (p2) -- (k2);
		\draw[->] (p2) -- (k3);
	\end{scope}

	\begin{scope}[every node/.style={fill=white, font=\tiny}]
		\node[align=center, font=\bf, rotate=90] at (-\gwidth*1.5-0.1,0) {Params};
		\draw[dashed] (-\gwidth*1.5,-\gheight/2)--(\gwidth*1.3,-\gheight/2);
		\node[align=center, font=\bf, rotate=90] at (-\gwidth*1.5-0.1,-\gheight) {KPMs};
	\end{scope}

	\begin{scope}[every node/.style={font=\tiny\bf,
					fill=white
				}]
		\node[align=center, text=Paired-F, inner sep=1pt] at (0,-\gheight/2) {KPM\\ CONFLICT};
	\end{scope}

\end{tikzpicture}}
		\caption{\gls{kpm} conflict in the \acrshort{kpm} Graph $\gk$.}
		\label{fig:gk}
	\end{subfigure}
	\ifexttikz\tikzsetnextfilename{three_graphs_3}\fi
	\begin{subfigure}[t]{0.32\linewidth}
		\centering
		\scalebox{0.8}{\usepgfplotslibrary{patchplots,groupplots,colorbrewer}
\pgfplotsset{
	compat=newest,
	plot coordinates/math parser=false,
	every tick label/.style={font=\tiny},
	colormap/Paired
}

\begin{tikzpicture}[auto]

	\begin{scope}[nodes={circle,draw,fill opacity=0.7}]
		\node[fill=Paired-G!80] (p1) at (-\gwidth,0) {$p_1$};
		\node[fill=Paired-G!80] (p2) at (0, 0) {$p_2$};
		\node[fill=Paired-G!80] (p3) at (\gwidth, 0) {$p_3$};
		\node[fill=Paired-G!80] (p4) at (-\gwidth, -\gheight) {$p_1$};
		\node[fill=Paired-G!80, draw=Paired-F] (p5) at (0, -\gheight) {$p_2$};
		\node[fill=Paired-G!80] (p6) at (\gwidth, -\gheight) {$p_3$};
	\end{scope}

	\begin{scope}[>=stealth']
		\draw[->] (p1) -- (p4);
		\draw[->, draw=Paired-F] (p1) -- (p5);
		\draw[->, draw=Paired-F] (p2) -- (p5);
		\draw[->] (p3) -- (p6);
	\end{scope}

	\begin{scope}[every node/.style={fill=white, font=\tiny}]
		\node[align=center, font=\bf, rotate=90] at (-\gwidth*1.5-0.1,0) {Params};
		\draw[dashed] (-\gwidth*1.5,-\gheight/2)--(\gwidth*1.3,-\gheight/2);
		\node[align=center, font=\bf, rotate=90] at (-\gwidth*1.5-0.1,-\gheight) {Params};
	\end{scope}

	\begin{scope}[every node/.style={font=\tiny\bf,
					fill=white
				}]
		E\node[align=center, text=Paired-F, inner sep=1pt] at (0,-\gheight/2) {PARAMTER\\ CONFLICT};
	\end{scope}

\end{tikzpicture}}
		\caption{Parameter conflict in the Parameter Graph $\gp$.}
		\label{fig:gp}
	\end{subfigure}

	\caption{Examples of conflicts and graphs used in \name.}
	\label{fig:three_graphs}
\end{figure}
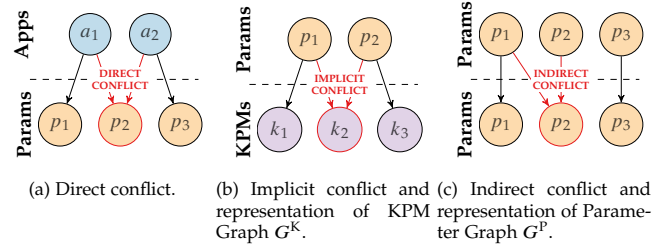

\begin{itemize}
	\item \textbf{Direct Conflicts} arise when applications control the same parameter (e.g., network slicing policies). An example of two applications controlling the same parameter $p_2$ is shown in Fig.~\ref{fig:direct}.
	      For example, this occurs when two xApps both control network slicing policies (potentially with conflicting objectives).
	\item \textbf{Indirect Conflicts}
	      occur when different applications control distinct parameters, but those parameters impact other parameters or \glspl{kpm} and the interdependencies can be observed.
	\item \textbf{Implicit Conflicts}
	      occur when different applications control distinct parameters, but those parameters impact other parameters or \glspl{kpm} and the interdependencies can not be observed.
\end{itemize}
\name's graph-based approach to the problem of conflicts in \oran makes it logical to further classify indirect and implicit conflicts into two categories. This classification is not based on the observability of the effect of an application on parameters and \glspl{kpm}, but rather on the structure of the graphs on which \name's analysis is based. Specifically, the two categories are:
\begin{itemize}
	\item \textbf{Parameter Conflicts}
	      occur when different applications control distinct parameters, but those parameters have interdependencies that cause undesired interactions.

	      An example is an rApp powering off a base station and an xApp adjusting its transmission power. An example of parameter $p_1$ indirectly affecting parameter $p_2$ is shown in Fig.~\ref{fig:gp}.
	\item \textbf{\gls{kpm} Conflicts}
	      arise when different applications control separate parameters that target different \kpm, but optimizing one metric can have unintended side effects on the metrics targeted by another application.

	      An example is that of an energy-saving rApp that jointly reduces bandwidth and transmission power

	      to improve energy-efficiency, while an xApp requests high resource utilization to support video streaming applications. Intuitively, both applications will impact the throughput experienced by served users. The former will negatively impact throughput due to the reduction in achievable capacity of the cell, while the latter will use many spectrum resources to deliver the highest throughput. An example of two parameters affecting the same \kpm $k_2$ is shown in Fig.~\ref{fig:gk}.
\end{itemize}

Implicit conflict detection assume no prior information on existence of conflicts, but this information can be built thanks to \gls{ml} techniques that build knowledge about conflicts from available data, such as the one proposed in~\cite{zolghadr2024learning}.

\subsection{\name's Conflict Model}\label{sec:conflict_model}

We now introduce \name's conflict model, as well as notation and graphs that will be used throughout the paper.

Let $\apps$ be the set of \oran applications that can be deployed on the \rics. For the sake of generality, in this work, we consider rApps, xApps and dApps, applications that extend \oran intelligence to the \glspl{cu} and \glspl{du}~\cite{d2022dapps}.
Let $\params$ be the set of parameters that can be controlled by applications in $\apps$, and $\metrics$ be the set of observable \kpms. For each application $a\in\apps$ and parameter $p\in\params$, we define an indicator $\alpha_{a,p}\in\{0,1\}$ such that $\alpha_{a,p}=1$ if application $a$ controls parameter $p$, and $\alpha_{a,p}=0$ otherwise. We can now define the set $\params^a=\{p\in\params : \alpha_{a,p}=1\}\subseteq \params$ to identify the subset of parameters that are controlled by $a$. An illustrative example with two applications $a_1$ and $a_2$, three parameters with $\params^{a_1}=\{p_1,p_2\}$ and $\params^{a_1}=\{p_2,p_3\}$ is shown in Fig.~\ref{fig:direct}.

\textbf{\kpm Graph $\gk$}: this graph $\gk=(V^{\mathrm{K}}, E^{\mathrm{K}})$ is shown in Fig.~\ref{fig:gk} and represents relationships between control parameters and \kpms. Nodes of $\gk$ are both control parameters and \kpms, i.e., $V^{\mathrm{K}} = \params \cup \metrics$, and edges $E^{\mathrm{K}}$ represent whether or not a control parameter $p\in\params$ impacts a certain \kpm $k\in\metrics$. Any 2-tuple $(p,k) \in \params \times \metrics$ is an edge of $\gk$, i.e., $(p,k) \in E^{\mathrm{K}}$ if and only if parameter $p$ directly affects \kpm $k$. Let $\epsilon_{p,k}\in\{0,1\}$ be an indicator variable such that $\epsilon_{p,k}=1$ if $p$ impacts \kpm $k$, and $\epsilon_{p,k}=0$ otherwise. We define $E^{\mathrm{K}} = \{ (p,m) \in \params \times \metrics : \epsilon_{p,k} = 1 \}$. In Fig.~\ref{fig:gk}, we show an example with two parameters (i.e., $\params=\{p_1,p_2\}$) and three \kpms (i.e., $\metrics=\{k_1,k_2,k_3\}$). $p_1$ impacts $k_1$ and $k_2$, while $p_2$ impacts $k_2$ and $k_3$. Since both $p_1$ and $p_2$ impact \acrshort{kpm} $k_2$, this causes a \gls{kpm} conflict.

\textbf{Parameter Graph $\gp$}: this graph is illustrated in Fig.~\ref{fig:gp} and is used to represent relationships among control parameters. Specifically, nodes of $\gp$ are $V^{\mathrm{P}} = \params$ and edges $E^{\mathrm{P}}$ are used to represent dependencies between parameters. For any 2-tuple $(p_1,p_2) \in \params \times \params$, let $\pi_{p_1,p_2}\in\{0,1\}$ be an indicator parameter such that $\pi_{p_1,p_2}=1$ if parameter $p_1$ impacts parameter $p_2$, $\pi_{p_1,p_2}=0$ otherwise. This graph aims at capturing dependencies between parameters and their respective conflicts, especially in those cases where the value of a parameter $p_1$ affects directly the value that $p_2$ can take as shown in Fig.~\ref{fig:gp}. For example, in the case that $p_1$ is used to turn off a base station, and $p_2$ represents its transmission power, then $p_2=0$ if the base station is off (i.e., $p_1=0$). Formally, we have $E^{\mathrm{P}} = \{ (p_1,p_2) \in \params \times \params : \pi_{p_1,p_2} = 1 \}$.

As we will discuss in Section~\ref{sec:detection}, the graph described above are used to identify relationships between applications, parameters and KPMs, and are used to determine which applications can potentially generate conflicts. However, in Section~\ref{sec:evaluation}, we show that severity of conflicts depends on operational conditions. Therefore, while the indicators and graphs described above show the potential of conflict occurrence, the relevance of such conflicts will be evaluated in each operational condition.

These graphs can be built using the concept of causality~\cite{nichols2007causal}, which allows to determine how a certain variable directly impacts the value of another variable. In this case, it makes it possible to determine direct causality relationships from observational data, and such relationships are represented via a directed graph~\cite{nichols2007causal}.

Given the \oran definitions of conflicts provided in Section~\ref{sec:conflictdefinition}, and the notations introduced before, we formally define these conflicts as follows. Since indirect and implicit conflicts only differ with respect to availability of knowledge, which can be built using \cite{zolghadr2024learning}, in the following we consider them together and focus on their parametric and \kpm sub-classification.

\begin{definition}[Direct Conflict]\label{def:direct}
	Let $p\in\params$ be any parameter. Let $a_1$ and $a_2$ be any two applications in $\apps$. We say that  $a_1$ and $a_2$ are in direct conflict with respect to parameter $p$ if $p \in \params^{\mathrm{DC}}_{a_1,a_2} = \params_{a_1} \cap \params_{a_2}$.
\end{definition}

\begin{definition}[Parameter Conflict]\label{def:paramconflict}
	Let $a_1$ and $a_2$ be any two applications in $\apps$. We say that parameter $p_1\in\params_{a_1}$ controlled by application $a_1$ generates a Parameter conflict with parameter $p_2\in\params_{a_2}$ controlled by $a_2$ if $(p_1,p_2) \in E^{\mathrm{P}}$, i.e., $\pi_{p_1,p_2} = 1$.
\end{definition}

\begin{definition}[\gls{kpm} Conflict]\label{def:kpmconflict}
	Let $a_1$ and $a_2$ be any two applications in $\apps$. We say that  $a_1$ and $a_2$ generate a \gls{kpm} conflict with respect to \kpm $k\in\metrics$ if there exists at least one 2-tuple $(p_1,p_2) \in \params_{a_1} \times \params_{a_2}$, with $p_1 \neq p_2$ such that $\epsilon_{p_1,k} = \epsilon_{p_2,k} = 1$.
\end{definition}

\vspace{-0.2cm}
\section{\name Overview} \label{sec:system}

\name has a modular architecture (Fig.~\ref{fig:system_architecture}) with four major logical blocks and a catalog of applications (e.g., rApps, xApps and dApps).
In this section, we will give a high-level overview of these blocks, while their detailed description will be given in Sections~\ref{sec:profiling}-\ref{sec:mitigation}.

\vspace{-0.2cm}
\subsection{\name in a Nutshell} \label{sec:pacifista_nutshell}

\name executes four major tasks, namely application profiling, conflict detection, evaluation, and mitigation.
The \textit{Profiler} runs on sandbox testing environments (e.g., digital twins, emulation environments) and generates profiles that describe the statistical behavior of \oran applications. This process will be detailed in Section~\ref{sec:stat_profiles}.
The \textit{Conflict Detection Module} uses such profiles for detecting the occurrence of conflicts and identifying the affected parameters and \kpms. Upon detecting the existence of conflicts, the \textit{Conflict Evaluation Module}, which is executed on the production network, generates a report that summarizes how severe the conflict is, and how much it impacts the \kpms. Finally, the \textit{Conflict Mitigation Module} leverages the information included in the report
to make informed decisions on the deployment of \oran applications. These decisions are made by using conflict management policies specified by the network operator, such as avoiding the deployment of an application that would generate too large of a conflict, or removing a subset of applications to reduce the severeness of conflicts below a certain threshold.

\vspace{-0.2cm}
\subsection{Integration with \oran}\label{sec:integration_with_oran}

As shown in Fig.~\ref{fig:system_architecture}, \name runs as a component of the \gls{smo}, and it leverages its internal messaging infrastructure to access the O1 termination and to interface with the \glspl{ric} and \gls{ran} nodes (e.g., \glspl{cu} and \glspl{du}). It is worth noticing that \name only needs to get access to application deployment and removal procedures. Following \oran specifications~\cite{polese2023understanding}, this only requires access to the O1 interface.
The profiling process, described in Section~\ref{sec:stat_profiles}, happens offline and consists of evaluating applications in one or more of the sandbox environments in Fig.~\ref{fig:system_architecture} (e.g., a digital twin or an isolated/test segment of the network). Each application is associated with a set of application profiles, one per operational condition.
Each operational condition specifies the wireless environment characteristics (e.g., channel conditions), traffic demand, mobility, and location of nodes that will be experienced by the application (e.g., xApp) upon deployment.
This information is not used by \name~for its computations, but it is
solely recorded to distinguish the different operational scenarios in which the
applications considered work.

First, \name gathers raw data and statistics on the decisions rApps, xApps, and dApps make based on the live \glspl{kpm} they get from the \ran nodes directly (dApps), through the E2 interface (xApps), and the O1 interface (rApps).
Then, \name performs profiling operations through statistical analysis on collected data by extracting \glspl{ecdf} for parameters and \kpms, and generates the application profiles to be included in the catalog.

As we will describe in Section~\ref{sec:mitigation}, the operator can specify conflict management policies that are used to determine which rApps, xApps, and dApps can be deployed on the \rics depending on the level of conflict they generate. Once \name makes a decision on the subset of \oran applications to deploy, the O1 interface is used to deploy xApps on the \nearrt \ric and dApps in RAN nodes, while the internal messaging infrastructure of the \gls{smo} is used to deploy rApps on the \nonrt \ric.
In both cases, applications are instantiated from a catalog hosted in the \gls{smo}.
Similarly, existing applications can be removed by \name in case they would conflict with new applications that need to be instantiated.
The O-RAN interfaces also enable \name to manage and monitor applications that have already been deployed.
As an example, the O1
and
the
R1 interfaces
can be used to perform health checks on the status of the running applications, or to tune their configuration.

\vspace{-0.2cm}
\section{Profiling \oran Applications}\label{sec:profiling}

The \textit{Catalog} hosts rApps, xApps, and dApps that can be deployed on an \oran network. For each application $a\in\apps$, the catalog stores an \textit{application profile} consisting of the following components:
\begin{itemize}
	\item \textbf{Identifier:} used to uniquely identify each application.
	\item \textbf{Parameter set:} this field specifies the list of parameters \textit{directly} controlled by the application (i.e., $\params_a$).
	\item \textbf{Statistical profile:} it provides statistical information on the behavior of the application under certain operational conditions. We consider the case where applications are profiled on a set $\ops$ of predefined operational conditions. Each operational condition $c\in\ops$ can specify, among others, the number of \glspl{ue}, cell load, \gls{sinr}/\gls{cqi} conditions. This information is available to operators via real-time and historical data. For each $c$, we store the statistical profile of the application which includes \glspl{pdf}, \glspl{cdf}, and \glspl{ecdf} used to characterize how application $a$ configures the parameters in $\params_a$ when operating under conditions $c$. In this paper, we use \glspl{ecdf} as these are model-free and can be extracted directly from data.
\end{itemize}

\begin{remark}
	To properly capture conflicts in \oran, it is important to notice that conflicts are strongly dependent on operational conditions. It is generally incorrect to state that two applications always generate conflicts. Indeed, two applications $a_1$ and $a_2$ might heavily conflict with each other under operational conditions $c_1$, but their conflict may be negligible under conditions $c_2$. For this reason, \name captures the statistical behavior of each application across $\ops$ and evaluates conflicts for each operational condition of interest.
	Moreover, the statistical profiling methodology employed by \name~is based on the assumption that there will be a certain level of noise and outliers. In order to account for these factors, the applications are executed for an extended period of time within the sandbox environment. This allows the collection of sufficient data to generate statistically valid results that include the noise and the possible outliers. It should be noted that outliers are not erroneous decisions; rather, they are decisions that do not follow the expected distribution. The presence of a significant number of outliers indicates that the generated profile does not align well with the curve of the real-world scenario, and it is necessary to create a new profile for that specific operational condition.
\end{remark}

\begin{remark}
	The parameter set is provided to the network operator by the application developer via a manifest. Among others, the manifest defines inputs and outputs of the algorithm implemented in the application, E2 service models required to run the model, controllable parameters and required \kpms.
\end{remark}

\vspace{-0.2cm}
\subsection{Creating Statistical Profiles} \label{sec:stat_profiles}

The creation of the statistical profiles is one of the most important aspects of \name as they allow it to detect and evaluate conflicts for mitigation and management. In \name, we generate statistical profiles by executing sandbox testing operations under each condition specified in $\ops$. In this context, ``sandbox" refers to an environment that simulates (or emulates) real-world network conditions and allows testing of \oran applications, network configurations, protocols, and services in a controlled virtual (e.g., digital twin) or physical (e.g., anechoic chamber) environment.

To generate accurate application profiles, it is necessary to collect enough samples to achieve statistical relevance for each operational condition, application and scenarios.
Indeed, statistical relevance varies across scenarios and applications. Therefore, the number of samples required to collect in each case is a design parameter that balances computational complexity with desired level of accuracy.
With respect to the applications considered in our experimental analysis (shown in Section~\ref{sec:experiments}), we collected at least 3,000 samples for each xApp for each slice.
Fig.~\ref{fig:statistical_relevance} shows how the application profile changes with the number of samples evaluated for slice \gls{mmtc} (similar results were obtained for other slices and xApps).
Fig.~\ref{fig:statistical_relevance_a5_ecdf_prb}~and \ref{fig:statistical_relevance_a5_ecdf_tp} show how the profiles for the variables considered are very similar, despite the vast range of samples analyzed.
Fig.~\ref{fig:statistical_relevance_a5_ks} shows how the \gls{ks} distance (more on this metric in Section \ref{sec:evaluation}) between each sample size and the total number of samples collected is very small ($< 0.02$ for the distance between the profile built with $3,000$ and $>60,000$ samples).
This means that the error on distance values due to having collected $\sim3,000$ samples instead of $>60,000$ is lower than $0.02$. Despite this being an acceptable error for the purpose of these experiments, we still used all the data available to build the profiles of each application.
Moreover, the accuracy of the application profiles depends on how accurate the sandbox testing environment is. The more accurate the sandbox environment (e.g., the digital twin), the more accurate the application profile. Finally, since applications behave differently under different operational conditions, \name generates an application profile for each operational condition to accurately capture conflicts under different deployments. In our implementation of \name (see Section~\ref{sec:experiments}), operational conditions have been specified by the number of base stations and their deployment location, the total number of \glspl{ue}, their mobility pattern, distribution, and traffic profiles.

\begin{figure}[t!]
	\ifexttikz
		\tikzsetnextfilename{statistical_relevance_a5_ks}
	\fi
	\begin{subfigure}{1\columnwidth}
		\centering
		\setlength\fwidth{1\columnwidth}
		\setlength\fheight{.18\columnwidth}
		\begin{tikzpicture}[]

	\begin{axis}[%
			width=0.85\fwidth,
			height=1\fheight,
			scale only axis,
			xmin=0.5,
			xmax=5.5,
			xtick={1,2,3,4,5},
			xticklabels={{500},{3000},{10000},{30000},{61851}},
			xlabel style={font=\scriptsize\color{white!15!black}},
			xlabel={N. samples},
			ymin=0,
			ymax=0.15,
			ytick={0,0.02,...,0.14},
			/pgf/number format/.cd,fixed,precision=2,fixed zerofill,
			ylabel style={font=\scriptsize\color{white!15!black}},
			ylabel={K-S distance},
			axis background/.style={fill=white},
			xmajorgrids,
			ymajorgrids,
			legend style={at={(0.97,0.97)}, anchor=north east, legend cell align=left, align=left, inner sep=0pt, draw=white!15!black, font=\scriptsize},
			reverse legend=false,
			ylabel shift=-3pt,
			xlabel shift=-4pt,
			enlargelimits=false
		]
		\addplot+ [
			color=Paired-F, mark=triangle, mark options={Paired-F},
			error bars/.cd, y dir=both, y explicit, error mark=-
		]
		table[row sep=crcr, y error plus index=2, y error minus index=3]{
				1 0.019840 0.071374 0.019840 \\
				2 0.008023 0.028556 0.008023 \\
				3 0.003992 0.013859 0.003992 \\
				4 0.001861 0.006595 0.001861 \\
				5 0.000000 0.000000 -0.000000 \\
			};
		\addlegendentry{Slice PRBs}

		\addplot [
			color=Paired-B, mark=o, mark options={solid, Paired-B},
			error bars/.cd, y dir=both, y explicit, error mark=-
		]
		table[row sep=crcr, y error plus index=2, y error minus index=3]{
				1 0.038058 0.103818 0.027701 \\
				2 0.015358 0.041753 0.011036 \\
				3 0.007921 0.021216 0.005374 \\
				4 0.003580 0.009727 0.002567 \\
				5 0.000000 0.000000 -0.000000 \\
			};
		\addlegendentry{Throughput (dl) [Mbps]}

	\end{axis}

\end{tikzpicture}%
		\vspace{-0.2cm}
		\caption{\gls{ks} distance among different samples sizes and the total number of samples collected for xApp $a_5$.}
		\label{fig:statistical_relevance_a5_ks}
	\end{subfigure}

	\ifexttikz
		\tikzsetnextfilename{statistical_relevance_a5_ecdf_prb}
	\fi
	\begin{subfigure}{1\columnwidth}
		\centering
		\setlength\fwidth{1\columnwidth}
		\setlength\fheight{.18\columnwidth}
		\begin{tikzpicture}[auto]

	\begin{axis}[%
			width=0.9\fwidth,
			height=1\fheight,
			scale only axis,
			xmin=0,
			xmax=33,
			xtick={0,3,...,36},
			xlabel style={font=\scriptsize\color{white!15!black}},
			xlabel={PRBs},
			ymin=0,
			ymax=1,
			ylabel style={font=\scriptsize\color{white!15!black}},
			ylabel={ECDF},
			axis background/.style={fill=white},
			title style={font=\scriptsize\color{black}},
			xmajorgrids,
			ymajorgrids,
			legend style={at={(0.03,0.97)}, anchor=north west, legend cell align=left, align=left, inner sep=0pt, draw=white!15!black, font=\tiny},
			legend columns=3,
			reverse legend=false,
			ylabel shift=-5pt,
			xlabel shift=-3pt,
			enlargelimits=false
		]
		\addplot[const plot, color=Paired-A, line width=1.0pt] table[row sep=crcr] {%
				21  0\\
				24  0.52\\
				27  0.986\\
				30  1\\
				30  1\\
			};
		\addlegendentry{N=500}

		\addplot[const plot, color=Paired-B, line width=1.0pt] table[row sep=crcr] {%
				18  0\\
				21  0.014\\
				24  0.52\\
				27  0.969\\
				30  1\\
				30  1\\
			};
		\addlegendentry{N=1000}

		\addplot[const plot, color=Paired-D, line width=1.0pt] table[row sep=crcr] {%
				21  0\\
				24  0.5067\\
				27  0.974\\
				30  1\\
				30  1\\
			};
		\addlegendentry{N=3000}

		\addplot[const plot, color=Paired-E, line width=1.0pt] table[row sep=crcr] {%
				21  0\\
				24  0.5097\\
				27  0.9753\\
				30  1\\
				30  1\\
			};
		\addlegendentry{N=7000}

		\addplot[const plot, color=Paired-F, line width=1.0pt] table[row sep=crcr] {%
				18  0\\
				21  0.0206\\
				24  0.5157\\
				27  0.9762\\
				30  1\\
				30  1\\
			};
		\addlegendentry{N=10000}

		\addplot[const plot, color=Paired-G, line width=1.0pt] table[row sep=crcr] {%
				18  0\\
				21  0.02147\\
				24  0.5175\\
				27  0.9782\\
				30  1\\
				30  1\\
			};
		\addlegendentry{N=30000}

		\addplot[const plot, color=Paired-H, line width=1.0pt] table[row sep=crcr] {%
				18  0\\
				21  0.02123\\
				24  0.5162\\
				27  0.9781\\
				30  1\\
				30  1\\
			};
		\addlegendentry{N=61851}

	\end{axis}

\end{tikzpicture}%
		\vspace{-0.5cm}
		\caption{\gls{ecdf} of assigned \glspl{prb} for different samples sizes for xApp $a_5$.}
		\label{fig:statistical_relevance_a5_ecdf_prb}
	\end{subfigure}

	\ifexttikz \tikzsetnextfilename{statistical_relevance_a5_ecdf_tp}
	\fi
	\begin{subfigure}{1\columnwidth}
		\centering
		\setlength\fwidth{1\columnwidth}
		\setlength\fheight{.15\columnwidth}
		\input{figurestikz/statistical_relevance_a5_ecdf_tp}
		\vspace{-0.5cm}
		\caption{\gls{ecdf} of downlink throughput for different samples sizes for xApp $a_5$.}
		\label{fig:statistical_relevance_a5_ecdf_tp}
	\end{subfigure}

	\vspace{-0.2cm}
	\caption{Application profile comparison for different number of samples for
		slice \gls{mmtc} of xApp $a_5$.}
	\label{fig:statistical_relevance}
\end{figure}

\begin{figure}[b!]
	\centering
	\includegraphics[width=\linewidth]{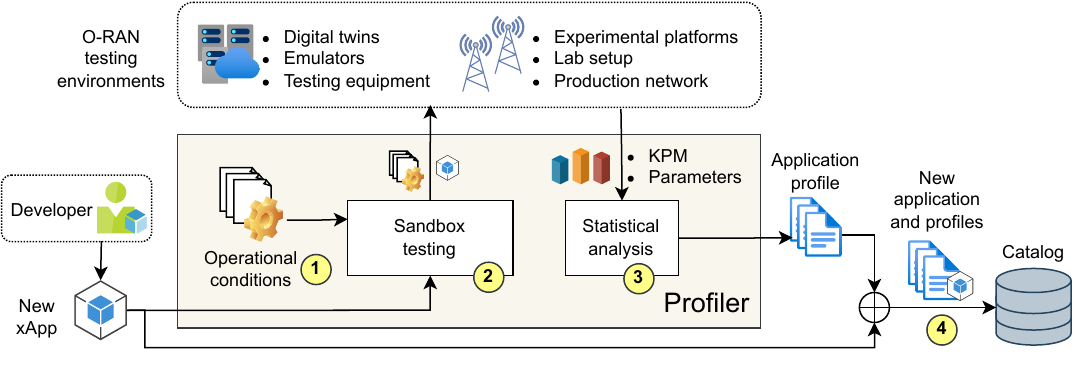}
	\vspace{-0.5cm}
	\caption{Profiling of new \oran applications.}
	\vspace{-0.1cm}
	\label{fig:profiling}
\end{figure}

In this work, we achieve this by using the Colosseum \oran digital twin~\cite{villa2024colosseum} and the OpenRAN Gym open-source \oran framework~\cite{bonati2023openran}. This procedure is illustrated in Fig.~\ref{fig:profiling} and described below.

\textbf{Step 1:}
This step consists in \textit{creating testing scenarios} to be included in $\ops$ and to be used as benchmarks to evaluate conflicts. These are generated according to the availability of the \oran testing environment. Digital twins,
network emulators, and testing equipment (e.g., \ric, \ran testers) are ideal platforms as these are controllable and reproducible environments. However, this does not exclude the use of over-the-air experimental platforms, lab setups, as well as portions of production networks. In this work, we leverage Colosseum to generate cellular scenarios by specifying topology (i.e., extracted from GPS coordinates from OpenCelliD~\cite{opencellid}), RF conditions (e.g., multi-path, fading), mobility, and traffic profiles, among others.
Details on these scenarios will be given in Section~\ref{sec:experiments}.
Note that an application $a\in\apps$ can only control $\params_a$, and those not controlled by $a$, i.e., $\params_{-a} = \params \setminus \params_a$, can assume multiple values. For this reason, we assume that each testing scenario $c$ also specifies the values of parameters in $\params_{-a}$. In this way, we can benchmark the same application $a$ across multiple testing scenarios that have the same topology, RF conditions, mobility and traffic profiles, but have different parameter configurations.

\textbf{Step 2:}
We select an application $a\in\apps$ and \textit{execute sandbox tests} under testing scenario $c\in\ops$. \name collects and logs data transmitted over O-RAN interfaces (e.g., O1, E2, A1) such as \kpms $\metrics$, control parameters $\params$ and enrichment information. We store both parameters controlled by $a$ and those that are not. Note that parameters in $\params_{-a}$ can be either fixed (e.g., assuming their default value), or change dynamically due to deterministic policies or due to other applications controlling them. In this latter case, where tests involve multiple applications executing at the same time (e.g., an xApp $a_1$ and an rApp $a_2$), we treat such applications as a ``virtual" application $a$ controlling $\params_a = \params_{a_1} \cup \params_{a_2}$.

\textbf{Step 3:} \name processes the data generated in the previous step to produce the statistical profile that describes the control behavior of application $a$ under testing scenario $c$, and the subsequent impact on \kpms.
In general, one could also store the original raw data in the statistical profile. However, this might be impractical due to its sheer size.\footnote{A single benchmark on our prototype generates $1.66$\:Mbps/\acrshort{gnb} of data when serving 6~\acrshortpl{ue} and storing more than 30 \kpms for each one of them.} For this reason, we have designed our conflict evaluation pipeline to only require statistical information of
data, i.e.,
\glspl{cdf} of \kpms, while raw data is stored in data lakes.

\textbf{Step 4:} Once statistical profiles have been generated, they are attached to application $a$ and published to the catalog.

\section{Detecting conflicts} \label{sec:detection}

\begin{figure}[t!]
	\centering
	\includegraphics[width=\linewidth]{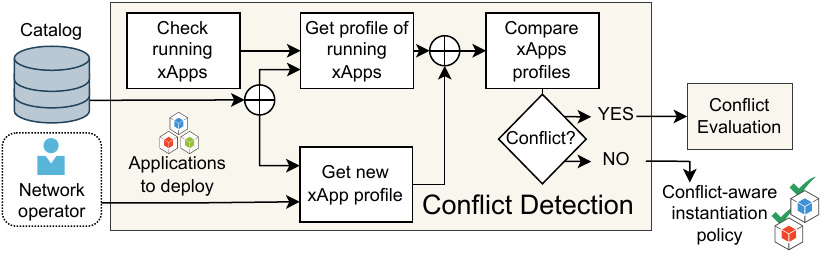}
	\vspace{-0.6cm}
	\caption{Conflict detection module and interactions with other modules.}
	\vspace{-0.6cm}
	\label{fig:module_detection}
\end{figure}

The first step in conflict mitigation consists in detecting the occurrence of conflicts. The \textit{Conflict Detection Module} takes as input the set of applications $\apps^*\subseteq\apps$ that the operator wants to deploy\footnote{In practical applications, this set can be represented as $\apps^* = \apps^\mathrm{new} \cup \apps^\mathrm{old}$, where $\apps^\mathrm{new}$ and $\apps^\mathrm{old}$ are the set of new applications that need to be deployed and the set of applications that are already deployed, respectively.} and (i) identifies the subset of applications that will generate conflicts; and (ii) identifies the set of parameters and \kpms that will be impacted. Fig.~\ref{fig:module_detection} shows how the Conflict Detection module works. Specifically, \name first extracts application profiles from the catalog, then compares the profiles to identify conflicts as described in the following sections. Ultimately, if no conflicts are detected, the applications are deployed directly. Otherwise, detected conflicts are sent to the conflict evaluation module described in Section~\ref{sec:evaluation}.

\vspace{-0.2cm}
\subsection{Detecting Direct Conflicts} \label{sec:detect:direct}

Following from Definition~\ref{def:direct}, the set $\params^{\mathrm{DC}}$ of parameters suffering from direct conflicts with respect to the application set $\apps^*$ is
\begin{align} \label{eq:dc_param}
	\confp^{\mathrm{DC}}(\apps^*) = \left\{p\in\bigcup_{a\in\apps^*} \params_a : \sum_{a\in\apps^*} \alpha_{a,p} > 1\right\}.
\end{align}
To identify these parameters, we augment the Parameter Graph $\gp$ by adding an extra layer above that represents the applications in $\apps^*$. Specifically, we add as many nodes as applications in $\apps^*$ and generate any edge $(a,p) \in \apps^*\times\params$ such that $\alpha_{a,p}=1$.

Fig.~\ref{fig:gp_extended} shows an example of this graph, which \name also uses to identify Parameter conflicts (see Section \ref{sec:detect:param}) with two applications and a total of three controllable parameters. Application $a_1$ controls $p_1$ and $p_2$, while $a_2$ controls $p_2$ and $p_3$. Fig.~\ref{fig:gp_extended} shows how the two applications generate a direct conflict with respect to $p_2$ as this parameter has more than one incoming edges.
For each parameter $p\in\confp^{\mathrm{DC}}(\apps^*)$, we also identify the subset of applications in $\apps^*$ that generate a direct conflict on $p$ as follows:
\begin{align}
	\confa^{\mathrm{DC}}_p(\apps^*) = \{a\in\apps^* : \alpha_{a,p} = 1\}.
\end{align}
In Fig.~\ref{fig:gp_extended}, we have that $\confp^{\mathrm{DC}}(\apps^*) = \{p_2\}$ and $\confa^{\mathrm{DC}}_{p_2}(\apps^*) = \{a_1, a_2\}$.

\vspace{-0.2cm}
\subsection{Detecting Parameter Conflicts} \label{sec:detect:param}

From Definition~\ref{def:paramconflict}, and by using the augmented graph built in Section~\ref{sec:detect:direct} and illustrated in Fig.~\ref{fig:gp_extended}, paramter conflicts can be characterized by identifying the following two sets:
\begin{align} \label{eq:idc_param}
	\confp^{\mathrm{PC}}(\apps^*)   & = \left\{p\in\bigcup_{a\in\apps^*} \params_a : \sum_{p'\in\bigcup_{a\in\apps^*}\params_a} \pi_{p',p} > 1\right\}, \\
	\confa^{\mathrm{PC}}_p(\apps^*) & = \{a\in\apps^* : \alpha_{a,p} = 1\}
\end{align}
\noindent
for each parameter $p\in\confp^{\mathrm{PC}}(\apps^*)$.
From Fig.~\ref{fig:gp_extended} (bottom part), we notice that parameter $p_3$ depends on $p_1$, i.e., $\pi_{p_1,p_3}=1$. Since $p_1\in\apps_{a_1}$, we have that decisions taken by $a_1$ inadvertently affect the value of parameters controlled by $a_2$, which is a Parameter conflict.

\ifexttikz
	\tikzsetnextfilename{gp_extended}
\fi
\begin{figure}[t!]
	\centering
	\scalebox{0.95}{\usepgfplotslibrary{patchplots,groupplots,colorbrewer}
\pgfplotsset{
	compat=newest,
	plot coordinates/math parser=false,
	every tick label/.style={font=\tiny},
	colormap/Paired
}

\begin{tikzpicture}[auto]

	\begin{scope}[nodes={circle,draw,fill opacity=0.7}]
		\node[fill=Paired-A] (a1) at (0,0) {$a_1$};
		\node[fill=Paired-A] (a2) at (2, 0) {$a_2$};
		\node[fill=Paired-G!80] (p1) at (-1, -1.5) {$p_1$};
		\node[fill=Paired-G!80, draw=Paired-F] (p2) at (1, -1.5) {$p_2$};
		\node[fill=Paired-G!80] (p3) at (3, -1.5) {$p_3$};
		\node[fill=Paired-G!80] (p4) at (-1, -3) {$p_1$};
		\node[fill=Paired-G!80] (p5) at (1, -3) {$p_2$};
		\node[fill=Paired-G!80, draw=Paired-F] (p6) at (3, -3) {$p_3$};
	\end{scope}

	\begin{scope}[>=stealth']
		\draw[->] (a1) -- (p1);
		\draw[->, draw=Paired-F] (a1) -- (p2);
		\draw[->, draw=Paired-F] (a2) -- (p2);
		\draw[->] (a2) -- (p3);
		\draw[->] (p1) -- (p4);
		\draw[->, draw=Paired-F] (p1) -- (p6);
		\draw[->] (p2) -- (p5);
		\draw[->, draw=Paired-F] (p3) -- (p6);
	\end{scope}

	\begin{scope}[every node/.style={fill=white}]
		\node[align=center, font=\bf] at (-2.5,0) {Apps};
		\draw[dashed] (-3,-0.75)--(3.5,-0.75);
		\node[align=center, font=\bf] at (-2.5,-1.5) {Params};
		\draw[dashed] (-3,-2.25)--(3.5,-2.25);
		\node[align=center, font=\bf] at (-2.5,-3) {Params};
	\end{scope}

	\begin{scope}[every node/.style={fill=white}]
		\footnotesize
		\node[align=center, text=Paired-F, font=\bf, inner sep=1.5pt] at (1,-0.75) {DIRECT\\ CONFLICT};
		\node[align=center, text=Paired-F, font=\bf, inner sep=1.5pt] at (2.3,-2.25) {PARAMETER\\ CONFLICT};
	\end{scope}

\end{tikzpicture}}
	\caption{Augmented graph $\gp$ to detect direct ($p_2$) and paramater ($p_3$) conflicts.}
	\vspace{-0.3cm}
	\label{fig:gp_extended}
\end{figure}

\subsection{Detecting \gls{kpm} Conflicts} \label{sec:detect:kpm}

\gls{kpm} conflicts are defined in Definition~\ref{def:kpmconflict}. Despite these conflicts being harder to model as they depend on intrinsic relationships between control parameters and observable \kpms, they can be detected using a procedure that is similar to that used for Parameter conflicts. Specifically, we augment the graph $\gk$ by adding nodes such that each node $a$ represents an application in $\apps^*$. Moreover, we also add edges $(a,p) \in \apps^* \times \params$ such that $\alpha_{a,p} = 1$.

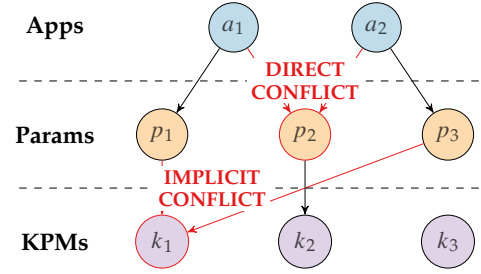
\begin{figure}[t!]
	\centering
	\scalebox{0.95}{\usepgfplotslibrary{patchplots,groupplots,colorbrewer}
\pgfplotsset{
	compat=newest,
	plot coordinates/math parser=false,
	every tick label/.style={font=\tiny},
	colormap/Paired
}

\begin{tikzpicture}[auto]

	\begin{scope}[nodes={circle,draw,fill opacity=0.7}]
		\node[fill=Paired-A] (a1) at (0,0) {$a_1$};
		\node[fill=Paired-A] (a2) at (2, 0) {$a_2$};
		\node[fill=Paired-G!80] (p1) at (-1, -1.5) {$p_1$};
		\node[fill=Paired-G!80, draw=Paired-F] (p2) at (1, -1.5) {$p_2$};
		\node[fill=Paired-G!80] (p3) at (3, -1.5) {$p_3$};
		\node[fill=Paired-I!80, draw=Paired-F] (k1) at (-1, -3) {$k_1$};
		\node[fill=Paired-I!80] (k2) at (1, -3) {$k_2$};
		\node[fill=Paired-I!80] (k3) at (3, -3) {$k_3$};
	\end{scope}

	\begin{scope}[>=stealth']
		\draw[->] (a1) -- (p1);
		\draw[->, draw=Paired-F] (a1) -- (p2);
		\draw[->, draw=Paired-F] (a2) -- (p2);
		\draw[->] (a2) -- (p3);
		\draw[->, draw=Paired-F] (p1) -- (k1);
		\draw[->] (p2) -- (k2);
		\draw[->, draw=Paired-F] (p3) -- (k1);
	\end{scope}

	\begin{scope}[every node/.style={fill=white}]
		\node[align=center, font=\bf] at (-2.5,0) {Apps};
		\draw[dashed] (-3,-0.75)--(3.5,-0.75);
		\node[align=center, font=\bf] at (-2.5,-1.5) {Params};
		\draw[dashed] (-3,-2.25)--(3.5,-2.25);
		\node[align=center, font=\bf] at (-2.5,-3) {KPMs};
	\end{scope}

	\begin{scope}[every node/.style={fill=white}]
		\footnotesize
		\node[align=center, text=Paired-F, font=\bf, inner sep=1.5pt] at (1,-0.75) {DIRECT\\ CONFLICT};
		\node[align=center, text=Paired-F, font=\bf, inner sep=1.5pt] at (-0.16,-2.25) {KPM\\ CONFLICT};
	\end{scope}

\end{tikzpicture}}
	\caption{Augmented graph $\gk$ to detect \gls{kpm} conflicts ($k_1$). The graph also shows a direct conflict at $p_2$.}
	\label{fig:gk_extended}
	\vspace{-0.5cm}
\end{figure}

Fig.~\ref{fig:gk_extended} shows an example of this graph. We can detect \gls{kpm} conflicts by identifying which \kpm nodes have more than one incoming edge. In this example, $a_1$ controls $\{p_1,p_2\}$, $a_2$ controls $\{p_2,p_3\}$. While the two applications generate a direct conflict on $p_2$, we notice that $k_1$ depends on both $p_1$ and $p_3$. As the two applications control both parameters, $k_1$ is affected by a parameter conflict.
\gls{kpm} conflicts can be identified via the following two sets:
\begin{align}
	\confp^{\mathrm{KC}}(\apps^*)   & = \left\{k\in\metrics : \sum_{p\in\bigcup_{a\in\apps^*}\params_a} \epsilon_{p,k} > 1\right\},  \label{eq:im_param} \\
	\confa^{\mathrm{KC}}_p(\apps^*) & = \{a\in\apps^* : \alpha_{a,p} = 1\} \label{eq:im_apps}
\end{align}
\noindent
for each parameter $p\in\confp^{\mathrm{KC}}(\apps^*)$.

\ifexttikz
	\tikzsetnextfilename{gk_extended}
\fi

\section{Evaluating conflicts} \label{sec:evaluation}

\begin{figure}[b]
	\setlength\abovecaptionskip{0pt}
	\centering
	\includegraphics[width=\linewidth]{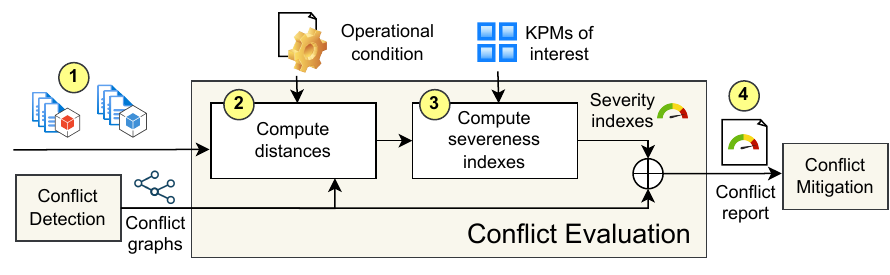}
	\vspace{0.05cm}
	\caption{Conflict evaluation module and interactions with other modules.}
	\label{fig:module_evaluation}
\end{figure}

Another important aspect of conflict management in \oran is that of conflict severity. This is particularly important as some conflicts might happen frequently, but their impact on network performance and efficiency might be tolerable under certain conditions.
The Conflict Evaluation Module in Fig.~\ref{fig:module_evaluation} analyzes each conflict detected in the previous phase over the application set $\apps^*$, and outputs a conflict report containing a set of indexes that measure the severity of conflicts and their potential impact on network performance. In our prototype, this procedure takes approximately twenty seconds.
In the following, we introduce a set of metrics that we use in \name to characterize conflicts, as well as methods to compute them via \name's Conflict Evaluation Module.

In Tab.~\ref{tab:distances}, we summarize the metrics
used by \name to provide an assessment of the severity of conflicts between \oran applications. To simplify the notation, we provide their general definition
for two one-dimensional generic \glspl{ecdf} $F_1(x)$ and $F_2(x)$ with $x\in\mathbb{R}$ being a random variable.
In our analysis, we have compared several distance metrics to identify the most suitable ones for the purpose of conflict evaluation. Specifically, we have selected \gls{ks} and \gls{int} for all numerical variables. As we will show in Section \ref{sec:distance-analysis}, the \gls{ks} distance is suitable for detecting conflicts, since the values of \gls{ks} distance are often close or equal to 1 in the case of conflict, and they are not as high otherwise. However, these \gls{ks} distances are not easily comparable as they only account for vertical distance between two \glspl{ecdf}, which does not provide insights on how different the two distributions are. On the contrary, the \gls{int} distance, which quantifies the area between two distributions, takes values that are more uniformly distributed in the $[0,1]$ interval. For this reason, the \gls{int} distance is convenient for making comparisons and measuring conflict severity.
Fig.~\ref{fig:distances_representations} shows a graphical representation of the two distances.
In case of categorical variables, \name uses the Pearson's Chi-Square test, where the distance between two applications is measured using the resulting $p$-value~\cite{hazra2016biostatistics}, as indicated in Tab.~\ref{tab:distances}. By definition, all distances we consider take values in $[0,1]$.

\begin{table}[t!]
	\setlength\abovecaptionskip{5pt}
	\setlength\belowcaptionskip{0pt}
	\centering
	\caption{Distance functions used in \name to evaluate conflicts in \oran.}
	\begin{tabular}[width=\textwidth]{c c p{0.6\linewidth}}
		\toprule
		\textbf{ID} & \textbf{Equation}                                     & \textbf{Description}                                                                              \\
		\midrule
		\gls{ks}    & $\max \lvert F_1(x) - F_2(x)\rvert$                   & Maximum vertical distance between the two ECDFs.                                                  \\
		\midrule
		\gls{int}   & $\sqrt{\frac{1}{L}\int \lvert F_1(x) - F_2(x)\rvert}$ & Integral of the absolute value of the distance between the two ECDFs, with $L=\max(x) - \min(x)$. \\
		\midrule
		$\chi$      & $1-\mathrm{p}$-value                                  & Likelihood that data from two categorical distributions are different.                            \\
		\bottomrule
	\end{tabular}
	\label{tab:distances}
\end{table}

\ifexttikz
	\tikzsetnextfilename{example_ks_int}
\fi
\begin{figure}[b!]
	\setlength\abovecaptionskip{-0pt}
	\setlength\belowcaptionskip{-0pt}
	\centering
	\setlength\fwidth{\columnwidth}
	\setlength\fheight{.2\columnwidth}
	\input{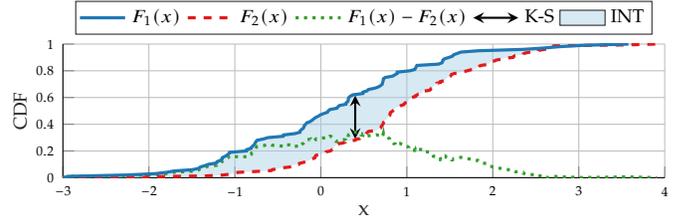}
	\caption{Graphical representation of the different distance metrics.}
	\label{fig:distances_representations}
\end{figure}

\name uses \glspl{ecdf}, which provide an accurate data-driven representation of the decision-making of each application for a certain operational condition $c\in\ops$.
The evaluation process in \name is executed as follows:

\textbf{Step 1:} We select two applications $a'$ and $a''$ and retrieve their statistical profile. We also select an operational condition $c\in\ops$ of interest. For each $p\in \params_{a'} \times \params_{a''}$, we extract the \glspl{ecdf} of the two applications with respect to $p$, say $F_{a'}(p|c)$ and $F_{a''}(p|c)$.

\textbf{Step 2:} We use Tab.~\ref{tab:distances} to compute the distance between $F_{a'}(p|c)$ and $F_{a''}(p|c)$ for each $p\in \params_{a'} \times \params_{a''}$. We refer to this distance as $D^f_{a',a''}(p|c)$, where $f$ represents the specific metric used to compute the distance as identified in Tab.~\ref{tab:distances}. For example, $D^{\mathrm{\gls{ks}}}_{a',a''}(p|c)$ represents the Kolmogorov-Smirnov (KS) distance between applications $a'$ and $a''$ with respect to control parameter $p$ under operational conditions $c$. Similarly, for each $k\in \metrics$, we extract the \glspl{ecdf} of $a'$ and $a''$ with respect to \kpm $k$. With a slight abuse of notation, we denote these \glspl{ecdf} as $F_{a'}(k|c)$ and $F_{a''}(k|c)$, respectively.

\textbf{Step 3:} We compute the distance between $F_{a'}(k|c)$ and $F_{a''}(k|c)$ for each $k\in\metrics$ via Tab.~\ref{tab:distances}. We use $D^f_{a',a''}(k|c)$ to indicate the distance between $a'$ and $a''$ with respect to \kpm $k$, under conditions $c$ and for a certain distance metric with identifier $f$ (from Tab.~\ref{tab:distances}).

\textbf{Step 4:} We combine the above distance metrics with respect to $f$ to generate two arrays $\mathbf{D}^f_{a',a''}(\params,c)=(D^f_{a',a''}(p|c))_{p\in\params}$ and $\mathbf{D}^f_{a',a''}(\metrics,c)=(D^f_{a',a''}(p|c))_{k\in\metrics}$. $\mathbf{D}^f_{a',a''}(\params,c)$ describes how applications $a'$ and $a''$ differ in terms of decision making policies (e.g., how differently they control the same set of parameters), while $\mathbf{D}^f_{a',a''}(\metrics,c)$ describes how the applications impact \kpms as a consequence of their different behavior. Both $\mathbf{D}^f_{a',a''}(\params,c)$ and $\mathbf{D}^f_{a',a''}(\metrics,c)$ are processed to generate a detailed report describing conflicts between $a'$ and $a''$. The report contains information regarding the existence of direct, parameter and \gls{kpm} conflicts, statistical information detailing how conflicts impact \kpms and parameters, as well as a set of indexes that are used by \name to express the severity of each conflict. The format of the report and the information contained therein will be described in Section~\ref{sec:report}.

\subsection{The Conflict Report} \label{sec:report}

\name leverages the information produced so far to generate the \textit{conflict report}. The objective of this report is twofold: (i) identify the existence of any type of conflict; and (ii) provide augmented information on how severe these conflicts are with respect to operators' objectives. Its generation is illustrated in Fig.~\ref{fig:module_evaluation}.

Operators can also specify the subset $\params^*$ and $\metrics^*$ of parameters and \kpms that are relevant to operator's goals and should be therefore considered when mitigating conflicts. For example, throughput might be an important \kpm for \gls{embb} applications, but be less relevant for \gls{urllc} traffic.
For a given set $\apps^*$ of applications to be evaluated, and sets $\params^*$ and $\metrics^*$, the report contains the following elements:
\begin{itemize}
	\item \textbf{Conflict Existence:}~the first elements included in the report are the sets $\confp^{\mathrm{DC}}(\apps^*)$, $\confp^{\mathrm{PC}}(\apps^*)$, $\confp^{\mathrm{KC}}(\apps^*)$, $\confa^{\mathrm{DC}}_p(\apps^*)$, $\confa^{\mathrm{PC}}_p(\apps^*)$, and $\confa^{\mathrm{KC}}_p(\apps^*)$ for each parameter $p\in\params$ as defined in \eqref{eq:dc_param}-\eqref{eq:im_apps}. These identify types of conflicts, which applications cause them and affected parameters and \kpms. \name also includes the augmented graphs (Section \ref{sec:detection}) used to visualize conflicts (Fig.~\ref{fig:gp_extended}).

	\item \textbf{Conflict Severity:}~for a given set $\apps^*$ of applications of interest with cardinality $A^*$, we have a total of $A^*(A^*-1)/2$ conflict pairs.\footnote{Note that the commutative property applies to conflict evaluation, i.e., evaluating the conflict for the pair $(a',a'')$ returns the same values as the evaluation for $(a'',a')$.} For each pair, \name computes two severity indexes $\sigma^{\mathrm{P}}_{a',a''}(\params^*|c)$ and $\sigma^{\mathrm{K}}_{a',a''}(\metrics^*|c)$. Each index summarizes how severe the different types of conflicts are by aggregating the distances $D^f_{a',a''}(z|c)$ computed in Tab.~\ref{tab:distances} for variable $z\in\params^*$ or $z\in\metrics^*$ under operational condition $c$ into a single value.
	      To combine the above distances and generate severity indexes for any given condition $c$ and distance metric $f$, we use a combining function $H(\cdot)$ such that $\sigma^{\mathrm{P}}_{a',a''}(\params^*|c) = H(\mathbf{D}^f_{a',a''}(\params^*,c))$ and $\sigma^{\mathrm{K}}_{a',a''}(\metrics^*|c) = H(\mathbf{D}^f_{a',a''}(\metrics^*,c))$.
	      Although $H(\cdot)$ can take any form, suitable aggregator functions for an $N$-dimensional array $\mathbf{x}=(x_1,\dots,x_N)$ are weighted average (i.e., $H(\mathbf{x})=1/N\sum_{i=1}^N w_i x_i$, where $w_i$ is the weight associated to each variable), median and maximum (i.e., $H(\mathbf{x})=\max\{x_1,\dots,x_N\}$) operators.
	      In particular, the weights $w_i$ of the average are used to weights the importance (or priority) of certain variables (\glspl{kpm} or parameters). For example, if the goal of the operator is to prioritize \gls{urllc} traffic over \gls{embb}, the weights in the aggregating function can be configured such that the weights associated to latency and reliability \glspl{kpm} of \gls{urllc} have larger values if compared to those related to \gls{embb}.

\end{itemize}

It is worth mentioning that \name computes the severity indexes based on the specific operational condition $c$. This is important as applications might generate conflicts only under certain conditions, and the severity of such conflicts might vary considerably under diverse operational conditions. For this reason, in \name operators need to specify the operational conditions of interest prior to generating a report.

\section{Mitigating conflicts} \label{sec:mitigation}

In practical deployments, conflicts can occur with non-zero probability due to coupling between control parameters and \kpms, limited amount of resources that result in competition between users, as well as conflicting intents (e.g., energy minimization against demand for high performance).
This means that operators either decide to deploy a handful of applications that can act in concert and serve a few types of subscribers to deliver specific services, or need to tolerate a certain degree of conflict. The first approach minimizes the occurrence of conflicts, but it also makes it difficult to satisfy performance requirements for a variety of services and applications. For this reason, we consider the second approach to showcase how the conflict evaluation of \name can be effectively used for more intelligent and dynamic conflict mitigation.

\name implements a threshold-based conflict mitigation strategy, where the tolerance level $\tol$ is used to identify applications that would generate too high a conflict and should not be deployed at that time. In \name, operators set a certain level of conflict tolerance $\tol$, which specifies the maximum level of conflict that the operator is willing to tolerate when deploying \oran applications. Since both \gls{ks} and \gls{int} assume values in $[0,1]$ and represent distances between \glspl{ecdf}, they are unitless. Therefore, $\tol\in [0,1]$ is also unitless.
The operator can also submit a priority index $I_a$ for each application $a\in\apps$, which reflects the importance the operator assigns to each application and is used by \name to determine which application to prioritize in the event of conflicts.
$\tol$ can be fixed for all deployment cycles, or set dynamically to a new value each time \name is run for a new group of applications.
$\tol$ can also be specified as an array of thresholds for each \gls{kpm} for each slice. Each entry is between 0 and 1, and is equal to 1 if the network operator does not have an interest in monitoring that specific \gls{kpm}. This allows the operator to have a more fine-grained control over the behavior of the network. In the following reasoning, it is assumed that the operator only provides a single threshold to be compared to the severity indices, which are a summary of the conflict level on the single \gls{kpm}.

\begin{figure}[t!]
	\centering
	\includegraphics[width=\linewidth]{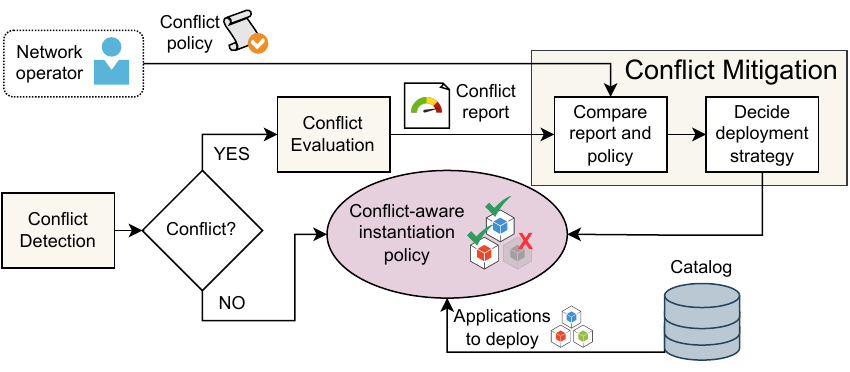}
	\vspace{-0.2cm}
	\caption{Conflict mitigation module and interactions with other modules.}
	\vspace{-0.6cm}
	\label{fig:module_mitigation}
\end{figure}

Upon receiving the set $\apps^*$ of applications to evaluate, the conflict mitigation module compares the severity indexes $\sigma^{\mathrm{P}}_{a',a''}(\params^*|c)$ and $\sigma^{\mathrm{K}}_{a',a''}(\metrics^*|c)$ (computed in Section~\ref{sec:report}) with the conflict threshold to determine which \oran applications to deploy to mitigate the occurrence of conflicts.
Specifically, since operators are more interested in tolerating conflicts with respect to their impact on \kpms rather than on parameter configurations, in the following we focus on $\sigma^{\mathrm{K}}_{a',a''}(\metrics^*|c)$.

The procedures involved in our threshold-based algorithm, shown in Fig.~\ref{fig:module_mitigation}, are as follows:

\textbf{Step 1:} \name identifies all the applications in $\apps^*$ that can be deployed without generating any type of conflict. These applications are added to a set $\apps^{\mathrm{DPLY}}$ of applications to deploy, which is initially set to $\apps^{\mathrm{DPLY}} = \emptyset$.

\textbf{Step 2:} We select the application with the highest priority from $\apps^*\setminus\apps^{\mathrm{DPLY}}$, i.e., $a = \arg\max_{a\in\apps^*\setminus\apps^{\mathrm{DPLY}}} \{I_a\}$. We add $a$ to $\apps^{\mathrm{DPLY}}$ if $\max_{a^*\in\apps^{\mathrm{DPLY}}}\{\sigma^{\mathrm{K}}_{a,a^*}(\metrics^*|c)\} \leq \tol$. We update $\apps^*=\apps^*\setminus{a}$.

\textbf{Step 3:} We repeat Step 2 until $\apps^*=\emptyset$.

The above procedure can also be extended to the onboarding of new applications, that is when deploying new applications when other applications have already been deployed. In this case, we set $\apps^{\mathrm{DPLY}} = \apps^\mathrm{old}$.

The selection of the value of threshold $\tol$ is indeed an important
aspect for the correct functioning of the system. It is strictly related to the
selection of the weights $w_i$ assigned to each variable for the weighted
average (as described in Section \ref{sec:report}).

\section{Experimental evaluation}\label{sec:experiments}

In this section, we first describe \name's prototype, and then present experimental results that illustrate how \name can be used to identify, characterize and mitigate conflicts in \oran.

\subsection{Prototype Description}
\name has been prototyped on the OpenRAN Gym framework~\cite{bonati2023openran} and tested experimentally on the Colosseum wireless network emulator~\cite{villa2024colosseum}, which enables at-scale experimentation with \glspl{sdr} as well as with realistic and heterogeneous \gls{rf} scenarios representative of real-world deployments.
Specifically, we leveraged the \scope and \coloran components of OpenRAN Gym~\cite{bonati2023openran} to instantiate a cellular network with a softwarized base station and 6 \glspl{ue}, and to deploy \nearrt \gls{ric} and xApps that interface with the base station through the O-RAN E2 termination. We use this as a sandbox environment to test a diverse set of xApps, collect data on their decision-making, and create statistical profiles used by \name
to verify the occurrence and severity of conflicts.

\ifexttikz
	\tikzsetnextfilename{stepcurves_slice_13_div_alt}
\fi
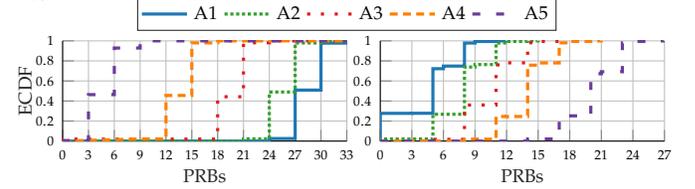
\begin{figure}[t!]
	\centering
	\setlength\abovecaptionskip{-0pt}
	\setlength\belowcaptionskip{-0pt}
	\setlength\fwidth{\columnwidth}
	\setlength\fheight{0.15\columnwidth}
	\usepgfplotslibrary{patchplots,groupplots,colorbrewer}
\pgfplotsset{
	compat=newest,
	plot coordinates/math parser=false,
	every tick label/.style={font=\tiny},
	colormap/Paired
}

\begin{tikzpicture}[font=\tiny]

	\begin{axis}[
			width=0.85\fwidth/2,
			height=\fheight,
			at={(0,0)},
			scale only axis,
			xmin=0,
			xmax=33,
			xtick={0,3,6,9,12,15,18,21,24,27,30,33,36},
			xlabel style={font=\scriptsize\color{white!15!black}},
			xlabel={PRBs},
			ymin=0,
			ymax=1,
			ylabel style={font=\scriptsize\color{white!15!black}},
			ylabel={ECDF},
			axis background/.style={fill=white},
			axis x line*=bottom,
			axis y line*=left,
			xmajorgrids,
			ymajorgrids,
			enlargelimits=false,
			legend style={inner sep=1pt, legend cell align=left,align=left,draw=white!15!black,anchor=south,font=\scriptsize,at={(0.85\fwidth/2, 1.1\fheight)}},
			legend columns=5,
			reverse legend=false,
			ylabel shift=-5pt,
			xlabel shift=-3pt,
			clip mode=individual  %
		]

		\addplot [color=Paired-B, very thick, solid]
		table[row sep=crcr]{%
				0   0\\
				24  0\\
				24	0.0251\\
				27	0.0251\\
				27	0.5089\\
				30	0.5089\\
				30	0.9777\\
				33	0.9777\\
				33	1\\
			};
		\addlegendentry{$a_1$}

		\addplot [color=Paired-D, very thick, densely dotted]
		table[row sep=crcr]{%
				0	  0\\
				18  0\\
				18	0.0001421\\
				21	0.0001421\\
				21	0.02245\\
				24	0.02245\\
				24	0.4897\\
				27	0.4897\\
				27	0.979\\
				30	0.979\\
				30	1\\
				33	1\\
			};
		\addlegendentry{$a_2$}

		\addplot [color=Paired-F, very thick, loosely dotted]
		table[row sep=crcr]{%
				0	  0\\
				0	  0.01987\\
				15	0.01987\\
				18	0.01987\\
				18	0.4425\\
				21	0.4425\\
				21	0.981\\
				24	0.981\\
				24	0.9998\\
				27	0.9998\\
				33	1\\
			};
		\addlegendentry{$a_3$}

		\addplot [color=Paired-H, very thick, densely dashed]
		table[row sep=crcr]{%
				0 	0\\
				9 	0.02094\\
				12	0.02094\\
				12	0.4553\\
				15	0.4553\\
				15	0.9798\\
				18	0.9798\\
				18	1\\
				33	1\\
			};
		\addlegendentry{$a_4$}

		\addplot [color=Paired-J, very thick, loosely dashed]
		table[row sep=crcr]{%
				0	  0\\
				0	  0.005384\\
				3	  0.005384\\
				3	  0.4635\\
				6	  0.4635\\
				6	  0.9286\\
				9	  0.9286\\
				9	  1\\
				33	1\\
			};
		\addlegendentry{$a_5$}

	\end{axis}

	\begin{axis}[
			width=0.85\fwidth/2,
			height=\fheight,
			at={(0.85\fwidth/2 + 0.1\fwidth/2,0)},
			scale only axis,
			xmin=0,
			xmax=27,
			xtick={0,3,6,9,12,15,18,21,24,27,30,33,36},
			xlabel style={font=\scriptsize\color{white!15!black}},
			xlabel={PRBs},
			ymin=0,
			ymax=1,
			ylabel style={font=\scriptsize\color{white!15!black}},
			axis background/.style={fill=white},
			axis x line*=bottom,
			axis y line*=left,
			xmajorgrids,
			ymajorgrids,
			enlargelimits=false,
			ylabel shift=-5pt,
			xlabel shift=-3pt,
			clip mode=individual  %
		]

		\addplot [color=Paired-B, very thick, solid]
		table[row sep=crcr]{%
				0	  0\\
				0	  0.2768\\
				2	  0.2768\\
				3	  0.2768\\
				3	  0.2776\\
				5	  0.2776\\
				5	  0.7223\\
				6	  0.7223\\
				6	  0.7485\\
				8	  0.7485\\
				8	  0.9785\\
				9	  0.9785\\
				9	  0.9939\\
				11	0.9939\\
				11	0.9995\\
				12	0.9995\\
				12	1\\
			};

		\addplot [color=Paired-D, very thick, densely dotted]
		table[row sep=crcr]{%
				0	  0\\
				0	  0.02093\\
				2	  0.02093\\
				5	  0.02093\\
				5	  0.2674\\
				6	  0.2674\\
				8	  0.2677\\
				8	  0.739\\
				9	  0.739\\
				9	  0.7638\\
				11	0.7638\\
				11	0.9793\\
				12	0.9793\\
				12	0.9946\\
				14	0.9946\\
				14	0.9999\\
				15	0.9999\\
				15	1\\
			};

		\addplot [color=Paired-F, very thick, loosely dotted]
		table[row sep=crcr]{%
				0	  0\\
				0	  0.0007967\\
				2	  0.0007967\\
				5	  0.0007967\\
				5	  0.01729\\
				8	  0.01729\\
				8	  0.3599\\
				9	  0.3599\\
				9	  0.3604\\
				11	0.3604\\
				11	0.7606\\
				12	0.7606\\
				12	0.7808\\
				14	0.7808\\
				14	0.9814\\
				15	0.9814\\
				15	0.9946\\
				17	0.9946\\
				17	1\\
			};

		\addplot [color=Paired-H, very thick, densely dashed]
		table[row sep=crcr]{%
				0	  0\\
				5	  0\\
				5	  0.0006873\\
				8	  0.0006873\\
				8	  0.01902\\
				11	0.01902\\
				11	0.2452\\
				12	0.2452\\
				12	0.2456\\
				14	0.2456\\
				14	0.7576\\
				15	0.7576\\
				15	0.7798\\
				17	0.7798\\
				17	0.9803\\
				18	0.9803\\
				18	0.9944\\
				20	0.9944\\
				20	0.9997\\
				21	0.9997\\
				21	1\\
			};

		\addplot [color=Paired-J, very thick, loosely dashed]
		table[row sep=crcr]{%
				0	  0\\
				11	0\\
				12	0.0002214\\
				14	0.0002214\\
				14	0.02098\\
				17	0.02098\\
				17	0.2519\\
				20	0.2519\\
				20	0.6717\\
				21	0.6717\\
				21	0.6923\\
				23	0.6923\\
				23	0.9797\\
				24	0.9797\\
				24	0.9942\\
				26	0.9942\\
				26	0.9997\\
				27	0.9997\\
				27	1\\
			};

	\end{axis}

\end{tikzpicture}
	\caption{\glspl{ecdf} of \gls{embb} (left) and \gls{urllc} (right) \gls{prb} allocation for $a_1$-$a_5$.}
	\label{fig:stepcurves_gaussians}
\end{figure}

To fairly compare all xApps against the same repeatable operational conditions $c$, we benchmark them on the Rome Colosseum scenario~\cite{bonati2023openran}---which reproduces the real-world cellular deployment in a section of Rome, Italy.
We consider 6~\glspl{ue} uniformly distributed within $50$\:m from the \gls{bs}, which allocates them in three network slices (\gls{embb}, \gls{urllc}, and \gls{mmtc}) across $10$\:MHz of spectrum (50~\glspl{prb} grouped into 17~\glspl{rbg}).
We leveraged the \gls{mgen}~\cite{mgen} tool to serve downlink traffic to the slice \glspl{ue} as follows: (i)~\gls{embb} \glspl{ue} are served constant bit rate traffic at a rate of $4$\:Mbps; (ii)~\gls{urllc} are served Poisson-distributed traffic at an average rate of $89.29$\:kbps; and (iii)~\gls{mmtc} \glspl{ue} are served Poisson-distributed traffic at an average rate of $44.64$\:kbps.

Our xApps act on a combination of two control parameters: (i)~the \textit{network slicing policy}, by adjusting the number of \glspl{prb} allocated to each slice; and (ii)~the \textit{scheduling policy} used in downlink transmissions, chosen among \gls{wf}, \gls{rr}, and \gls{pr}.
Although \name is general and can be used to profile any \oran application that embeds logic to control \ran parameters, in our experimental evaluation we focus on two classes of xApps: stochastic and \gls{drl}-based xApps. Stochastic xApps embed predictable decision-making logic that will be used to showcase \name functionalities, and highlight the importance of identifying and characterizing conflicts. \gls{drl}-based xApps include a selection of xApps taken from the literature~\cite{tsampazi2023globecom} and embed \gls{drl} agents trained to satisfy intents and diverse slice requirements.

\subsubsection{Stochastic xApps ($a_1$-$a_5$)} \label{sec:xapps:deterministic}

This set includes 5 xApps controlling slicing policies generated via a Gaussian distribution with different mean values per slice and a standard deviation of 1.5.
xApps execute in real time and generate a new random slicing policy every $250$~ms based on the distribution assigned to the xApp. Each new \gls{prb} assignment policy is generated by drawing a value for each slice around the mean value of the \gls{prb} distribution for the slice with a variance of 1.5. Prior to transmission, each random draw is rounded to the closest \glspl{rbg} allocation, and then the policy is checked to avoid allocating more than the 50 available \glspl{prb}.
\glspl{ecdf} for xApps $a_1$-$a_5$ and slices \gls{embb} and \gls{urllc} are shown in Fig.~\ref{fig:stepcurves_gaussians}. We omit the \gls{ecdf} for slice \gls{mmtc}, which receives the remaining \glspl{prb}.

\subsubsection{\gls{drl}-based xApps ($a_6$-$a_8$)}\label{sec:xapps:drl}
These xApps are taken from \cite{tsampazi2023globecom} and embed \gls{drl} agents controlling a combination of slicing (i.e., the portion of the available \glspl{prb} allocated to each slice) and scheduling (i.e., how the \glspl{prb} are internally allocated to users of each slice) policies to satisfy slice-specific intents. Specifically, they all aim at maximizing \gls{embb} throughput and number of \gls{mmtc} transmitted packets, while minimizing \gls{dl} buffer size for \gls{urllc} traffic (as a proxy of latency). They all target this goal via different action spaces. $a_6$ controls the scheduling policy only, $a_7$ controls the slicing policy only, while $a_8$ controls both the scheduling and the slicing policies. These are referred to as \texttt{\small Sched 0.5}, \texttt{\small Slicing 0.5}, \texttt{\small Sched \& Slicing 0.5} in \cite{tsampazi2023globecom}, respectively.
These xApps embed agents trained using \gls{ppo}, a state-of-the-art \gls{rl} architecture~\cite{chen2023static}, and are deployed inside the Near-RT \gls{ric}. They receive real-time \glspl{kpm} via the E2 interface, make decisions based on network conditions such as downlink throughput, buffer occupancy, and the number of transmitted packets, and continuously adapt their control policies based on real-time feedback from the \gls{ran}. It is noted that the model-free architecture and trial-and-error approach of the \gls{ppo} algorithm are an ideal fit for enhancing the resource allocation process in stochastic environments, such as wireless channels. Ultimately, our goal is to demonstrate that xApps with similar intents,
even when controlling different parameters, are prone to generating minimal conflicts.

\begin{table}[t!]
	\arrayrulecolor{white} %
	\centering
	\setlength\abovecaptionskip{5pt}
	\setlength\belowcaptionskip{5pt}
	\caption{\gls{ks} (Left) and \gls{int} (Right) distances for \gls{embb} with respect to xApp $a_1$. The last row shows the severity index for the corresponding xApp comparison.}
	\begin{tabular}{c|*{3}{E} E c *{4}{E|}}
		\cline{1-8}
		                               & \multicolumn{1}{c}{$D^{\mathrm{K-S}}_{1,2}$} & \multicolumn{1}{c}{$D^{\mathrm{K-S}}_{1,3}$} & \multicolumn{1}{c}{$D^{\mathrm{K-S}}_{1,4}$} & \multicolumn{1}{c}{$D^{\mathrm{K-S}}_{1,5}$} &  & \multicolumn{1}{c}{$D^{\mathrm{INT}}_{1,2}$} & \multicolumn{1}{c}{$D^{\mathrm{INT}}_{1,3}$} & \multicolumn{1}{c}{$D^{\mathrm{INT}}_{1,4}$} & \multicolumn{1}{c}{$D^{\mathrm{INT}}_{1,5}$} \\
		\hhline{~*8{|-}|}
		PRBs                           & 0.47                                         & 0.98                                         & 1.00                                         & 1.00                                         &  & 0.29                                         & 0.49                                         & 0.64                                         & 0.81                                         \\
		\hhline{~*8{|-}|}
		Buffer Size                    & 0.15                                         & 0.32                                         & 0.43                                         & 0.85                                         &  & 0.29                                         & 0.49                                         & 0.45                                         & 0.78                                         \\
		\hhline{~*8{|-}|}
		Throughput                     & 0.13                                         & 0.29                                         & 0.36                                         & 0.82                                         &  & 0.13                                         & 0.22                                         & 0.21                                         & 0.41                                         \\
		\hhline{-*8{|-}|}
		\hhline{-*8{|-}|}
		\arrayrulecolor{black} %
		\arrayrulecolor{white} %
		\hhline{-*8{|-}|}
		\hhline{-*8{|-}|}
		\hhline{-*8{|-}|}
		\hhline{-*8{|-}|}
		\hhline{-*8{|-}|}
		\hhline{-*8{|-}|}
		Severity $\sigma^{\mathrm{K}}$ & 0.14                                         & 0.30                                         & 0.40                                         & 0.84                                         &  & 0.21                                         & 0.35                                         & 0.33                                         & 0.59                                         \\
		\hhline{~*8{|-}|}
		\arrayrulecolor{black} %
	\end{tabular}
	\label{tab:combined_distances}
\end{table}

\ifexttikz
	\tikzsetnextfilename{int_over_ks_slice1_prbs}
\fi
\begin{figure}[b!]
	\setlength\abovecaptionskip{-0pt}
	\setlength\belowcaptionskip{-0pt}
	\centering
	\setlength\fwidth{\columnwidth}
	\setlength\fheight{.25\columnwidth}
	\usepgfplotslibrary{patchplots,groupplots,colorbrewer}
\pgfplotsset{
	compat=newest,
	plot coordinates/math parser=false,
	every tick label/.style={font=\tiny},
	colormap/Paired
}

\begin{tikzpicture}[font=\tiny]

	\begin{axis}[
			width=0.9\fwidth,
			height=\fheight,
			scale only axis,
			xmin=0,
			xmax=36,
			xtick={0,3,6,9,12,15,18,21,24,27,30,33,36},
			xlabel style={font=\scriptsize\color{white!15!black}},
			xlabel={PRBs},
			ymin=0,
			ymax=1,
			ylabel style={font=\scriptsize\color{white!15!black}},
			ylabel={ECDF},
			axis background/.style={fill=white},
			axis x line*=bottom,
			axis y line*=left,
			xmajorgrids,
			ymajorgrids,
			enlargelimits=false,
			legend style={inner sep=1pt, legend cell align=left,align=left,draw=white!15!black,anchor=south,font=\scriptsize,at={(0.5,1.07)}},
			legend columns=6,
			reverse legend=true,
			ylabel shift=-3pt,
			xlabel shift=-3pt,
			clip mode=individual  %
		]

		\addplot[area legend, draw=none, fill=Paired-F, fill opacity=0.35]
		table[row sep=crcr] {
				x	y\\
				0	  0\\
				0	  0.01987\\
				15	0.01987\\
				18	0.01987\\
				18	0.4425\\
				21	0.4425\\
				21	0.981\\
				24	0.981\\
				24	0.9998\\
				27	0.9998\\
				33	1\\
				33	0.9777\\
				30	0.9777\\
				30	0.5089\\
				27	0.5089\\
				27	0.0251\\
				24	0.0251\\
				24	0\\
			}--cycle;
		\addlegendentry{$D^{\text{INT}}_{1,3}$}

		\addplot[area legend, draw=none, pattern=north east lines, pattern color=Paired-B]
		table[row sep=crcr] {
				x	y\\
				0	  0\\
				0	  0.005384\\
				3	  0.005384\\
				3	  0.4635\\
				6	  0.4635\\
				6	  0.9286\\
				9	  0.9286\\
				9	  1\\
				33	1\\
				33	0.9777\\
				30	0.9777\\
				30	0.5089\\
				27	0.5089\\
				27	0.0251\\
				24	0.0251\\
				24	0\\
			}--cycle;
		\addlegendentry{$D^{\text{INT}}_{1,5}$}

		\addplot [color=black, thick,
		{stealth[length=2mm,inset=0pt]}-{stealth[length=2mm,inset=0pt]}]
		table[row sep=crcr]{
				21	0\\
				21	0.981\\
			};
		\addlegendentry{$D^{\text{K-S}}$}

		\addplot [color=Paired-B, very thick, densely dotted]
		table[row sep=crcr]{
				0	  0\\
				0	  0.005384\\
				3	  0.005384\\
				3	  0.4635\\
				6	  0.4635\\
				6	  0.9286\\
				9	  0.9286\\
				9	  1\\
				33	1\\
			};
		\addlegendentry{$a_5$}

		\addplot [color=Paired-F, very thick, dashed]
		table[row sep=crcr]{
				0	  0\\
				0	  0.01987\\
				15	0.01987\\
				18	0.01987\\
				18	0.4425\\
				21	0.4425\\
				21	0.981\\
				24	0.981\\
				24	0.9998\\
				27	0.9998\\
				33	1\\
			};
		\addlegendentry{$a_3$}

		\addplot [color=Paired-D, very thick]
		table[row sep=crcr]{
				0   0\\
				24  0\\
				24	0.0251\\
				27	0.0251\\
				27	0.5089\\
				30	0.5089\\
				30	0.9777\\
				33	0.9777\\
				33	1\\
			};
		\addlegendentry{$a_1$}

		\addplot [color=black, thick,
		{stealth[length=2mm,inset=0pt]}-{stealth[length=2mm,inset=0pt]}, forget plot]
		table[row sep=crcr]{
				21	0\\
				21	0.981\\
			};
		\addplot [color=black, thick,
		{stealth[length=2mm,inset=0pt]}-{stealth[length=2mm,inset=0pt]}, forget plot]
		table[row sep=crcr]{
				9	0\\
				9	1\\
			};

		\begin{scope}[every node/.style={font=\scriptsize,fill=white,draw=black,inner sep=1pt}]
			\node [rotate=90, anchor=center, draw=Paired-B] at (axis cs: 10.5, 0.5) {$D^{\text{K-S}}_{1,5}\!=\!1.00$};
			\node [rotate=90, anchor=center, draw=Paired-F] at (axis cs: 22.5, 0.5) {$D^{\text{K-S}}_{1,3}\!=\!0.98$};
			\node [anchor=west, draw=Paired-B] at (axis cs: 12.5, 0.8) {$D^{\text{INT}}_{1,5}\!=\!0.81$};
			\node [anchor=west, draw=Paired-F] at (axis cs: 24, 0.8) {$D^{\text{INT}}_{1,3}\!=\!0.49$};
		\end{scope}

	\end{axis}

\end{tikzpicture}
	\caption{\glspl{ecdf} of the number of \glspl{prb} assigned to the \gls{embb} slice of xApps $a_1$, $a_3$, and $a_5$ and corresponding \gls{ks} and \gls{int} distances.}
	\label{fig:int_over_ks_slice1_prbs}
\end{figure}
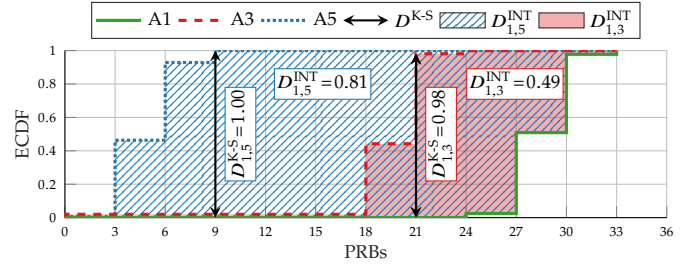

\subsection{Experimental Results}\label{sec:distance-analysis}

In our experiments, we consider both \gls{ks} and \gls{int} distances (shown in Tab.~\ref{tab:distances}).
The conflict analysis and report generation are performed by the Conflict Detection and Conflict Evaluation modules. In our prototype, both modules are implemented in MATLAB.
In the following, distances between control parameters refer to direct conflicts, while distances between \kpms refer to \gls{kpm} conflicts. In our analysis, the set of parameters and \kpms considered consists of \glspl{prb}, Scheduling Policy, Buffer Size, and Throughput, since these show correlation with all the other \kpms that have been collected to assess the network performance. In the following section, results are shown for slices \gls{embb} and \gls{urllc} only, excluding slice \gls{mmtc} because slice \gls{urllc} and \gls{mmtc} have similar characteristics and show similar results. Moreover, although we only consider downlink \kpms, \name is agnostic to the physical meaning of \kpms and only focuses on how actions taken by any \oran application impact the value of such \kpms.

\subsubsection{Relevance of Different Distance Functions}\label{sec:different_distance_functions}
We first illustrate how the \gls{ks} distance is a good indicator for detecting conflicts, while the \gls{int} distance brings more granular insights on conflict severity.
This is shown in Tab.~\ref{tab:combined_distances}, where we compare \gls{ks} and \gls{int} distances taking xApp $a_1$ as reference.
In general, we notice that \gls{ks} distances are larger than \gls{int}. This is better illustrated in Fig.~\ref{fig:int_over_ks_slice1_prbs}, where we notice that the \gls{ks} distances between $a_1$ and $a_3$, and $a_1$ and $a_5$ are both close to 1, thus showing the existence of direct conflict. However, the \gls{int} distances in the two cases are 0.49 and 0.81, respectively. This shows that although there is direct conflict in both cases, this is less severe in the $a_1$-$a_3$ case than in the $a_1$-$a_5$ case as $a_1$ and $a_3$ compute similar slicing policies. The last row in Tab.~\ref{tab:combined_distances} also shows conflict severity measured using an average combining function $H(\cdot)$ with $\metrics^*$ containing both downlink throughput and buffer size. As expected, conflict severity increases with \kpm distances which, in this case, are averaged via $H(\cdot)$. Severity indexes calculated using \gls{int} distances for all stochastic xApps are also reported in Tab.~\ref{tab:distances_apps_a3_2_a7_embb_int_averaged} and \ref{tab:distances_apps_a3_2_a7_urllc_int_averaged} for \gls{embb} and \gls{urllc} slices, respectively.

\label{fig:int_over_ks_slice3_buffersize-zoom}

\begin{table}[b!]
	\centering
	\begin{minipage}[t]{0.49\linewidth}
		\caption{Severity indexes $\sigma^{\mathrm{K}}$ using \gls{int} distance for \gls{embb}.}
		\begin{tabular}{c*{5}{F|} F}
			                          & \multicolumn{1}{c}{$a_1$} & \multicolumn{1}{c}{$a_2$} & \multicolumn{1}{c}{$a_3$} & \multicolumn{1}{c}{$a_4$} & \multicolumn{1}{c}{$a_5$} \\
			\hhline{~*6{|-}|}
			\multicolumn{1}{c}{$a_1$} & .00                       & .21                       & .35                       & .33                       & .59                       \\
			\hhline{~*6{|-}|}
			\multicolumn{1}{c}{$a_2$} & .21                       & .00                       & .28                       & .26                       & .55                       \\
			\hhline{~*6{|-}|}
			\multicolumn{1}{c}{$a_3$} & .35                       & .28                       & .00                       & .19                       & .48                       \\
			\hhline{~*6{|-}|}
			\multicolumn{1}{c}{$a_4$} & .33                       & .26                       & .19                       & .00                       & .00                       \\
			\hhline{~*6{|-}|}
			\multicolumn{1}{c}{$a_5$} & .59                       & .55                       & .48                       & .00                       & .00                       \\
			\hhline{~*6{|-}|}
		\end{tabular}
		\label{tab:distances_apps_a3_2_a7_embb_int_averaged}
	\end{minipage}\hfill
	\begin{minipage}[t]{0.49\linewidth}
		\centering
		\caption{Severity indexes $\sigma^{\mathrm{K}}$ using \gls{int} distance for \gls{urllc}.}
		\begin{tabular}{c*{6}{E|}}
			                          & \multicolumn{1}{c}{$a_1$} & \multicolumn{1}{c}{$a_2$} & \multicolumn{1}{c}{$a_3$} & \multicolumn{1}{c}{$a_4$} & \multicolumn{1}{c}{$a_5$} \\
			\hhline{~*6{|-}|}
			\multicolumn{1}{c}{$a_1$} & .0000                     & .0101                     & .0121                     & .0123                     & .0134                     \\
			\hhline{~*6{|-}|}
			\multicolumn{1}{c}{$a_2$} & .0101                     & .0000                     & .0066                     & .0071                     & .0088                     \\
			\hhline{~*6{|-}|}
			\multicolumn{1}{c}{$a_3$} & .0121                     & .0066                     & .0000                     & .0047                     & .0064                     \\
			\hhline{~*6{|-}|}
			\multicolumn{1}{c}{$a_4$} & .0123                     & .0071                     & .0047                     & .0000                     & .0000                     \\
			\hhline{~*6{|-}|}
			\multicolumn{1}{c}{$a_5$} & .0134                     & .0088                     & .0064                     & .0000                     & .0000                     \\
			\hhline{~*6{|-}|}
		\end{tabular}
		\label{tab:distances_apps_a3_2_a7_urllc_int_averaged}
	\end{minipage}
\end{table}

We notice that conflicts impact largely the \gls{embb} slice, where xApps that allocate less \glspl{prb} to \gls{embb} (e.g., $a_4$, $a_5$) severely degrade throughput (i.e., target \kpm for \gls{embb}). Instead, since \gls{urllc} requests less traffic, we notice that conflict severity is very small and close to 0 in general, i.e., \gls{urllc} is minimally affected by \gls{kpm} conflicts caused by xApps with high direct conflicts as the buffer is emptied even with the few \glspl{prb} allocated by $a_1$ to \gls{urllc}.

\subsubsection{DRL-Based xApps}
In the first experiment, we consider xApps $a_6$-$a_8$ which use \gls{drl} agents with diverse action spaces to improve slice-specific \kpms (Section~\ref{sec:xapps:drl}). Due to space limitations, we only present conflict analysis for \gls{embb} and \gls{urllc} slices. We report downlink throughput values for the former, and downlink buffer size for the latter.
Tab.~\ref{tab:distances_numbers_intent_based} shows the K-S and INT distances for numerical parameters (i.e., slicing policies) and \kpms, and $\chi$ distance for scheduling policies, as well as the impact of conflicts on relevant \kpms. Recall that $a_8$ controls both slicing and scheduling policies, while $a_6$ and $a_7$ respectively control scheduling and slicing only. Therefore, there is no direct conflict on scheduling between $a_8$ and $a_7$, and no direct conflict on slicing between $a_8$ and $a_6$.
Since $a_7$ and $a_8$ both control slicing policies, they generate a direct conflict with K-S and INT distance equal to 0.11 and 0.23, respectively. The same holds for $a_6$ and $a_8$ which produce a direct conflict with respect to scheduling with a $\chi$ distance of 0.61, suggesting that the two xApps select different scheduling policies. In general, conflicts have low values due to the shared goal. However, we notice that controlling slicing policies (i.e., $a_7$ and $a_8$) results in lower \gls{kpm} conflicts for both throughput and buffer size (i.e., the largest distance in this case is $D^{\mathrm{K-S}}_{8,7}=0.14$), suggesting that controlling slicing policies under a shared goal makes it possible to achieve higher performance than scheduling control alone~\cite{tsampazi2023globecom}, which allows the xApps to better satisfy the shared intent and produce less conflicts. We also notice that the larger action space (i.e., scheduling and slicing) allows $a_8$ to improve performance.
For example, Fig.~\ref{fig:drl_apps_int_over_ks_slice1_throughput} shows how $a_8$ delivers higher \gls{embb} throughput than $a_7$.

\begin{table}[b!]
	\addtolength{\tabcolsep}{-1pt} %
	\setlength\abovecaptionskip{5pt}
	\setlength\belowcaptionskip{5pt}
	\centering
	\caption{Direct and \gls{kpm} conflict analysis taking $a_8$ as the reference xApp.}
	\begin{tabular}{ c c *{6}{c} }
		\arrayrulecolor{black} %
		\toprule
		\textbf{Slice} & \textbf{Variable} & \multicolumn{3}{c}{\textbf{$a_8$-$a_6$}}    & \multicolumn{3}{c}{\textbf{$a_8$-$a_7$}}                                                                                                                                                                     \\
		\cmidrule(lr){3-5} \cmidrule(lr){6-8}
		               &                   & {$D^{\mathrm{K-S}}_{8,6}$}   \hspace{0.3cm} & {$D^{\mathrm{INT}}_{8,6}$} \hspace{0.3cm} & {$D^{\mathrm{\chi}}_{8,6}$} \hspace{0.3cm} & {$D^{\mathrm{K-S}}_{8,7}$} \hspace{0.3cm} & {$D^{\mathrm{INT}}_{8,7}$} \hspace{0.3cm} & {$D^{\mathrm{\chi}}_{8,7}$} \\ \midrule
		\gls{embb}     & {\glspl{prb}}     & 0                                           & 0                                         & $-$                                        & 0.11                                      & 0.23                                      & $-$                         \\
		\gls{embb}     & {Scheduling}      & $-$                                         & $-$                                       & 0.61                                       & $-$                                       & $-$                                       & 0                           \\
		\gls{embb}     & {Throughput}      & 0.46                                        & 0.27                                      & $-$                                        & 0.14                                      & 0.13                                      & $-$                         \\ \midrule
		\gls{urllc}    & {\glspl{prb}}     & 0                                           & 0                                         & $-$                                        & 0.10                                      & 0.16                                      & $-$                         \\
		\gls{urllc}    & {Scheduling}      & $-$                                         & $-$                                       & 0.19                                       & $-$                                       & $-$                                       & 0                           \\
		\gls{urllc}    & {Buffer Size}     & 0.06                                        & 0.01                                      & $-$                                        & 0.04                                      & 0.01                                      & $-$                         \\ \bottomrule
	\end{tabular}
	\label{tab:distances_numbers_intent_based}
\end{table}

\ifexttikz
	\tikzsetnextfilename{drl_apps_int_over_ks_slice1_throughput}
\fi
\begin{figure}[t!]
	\setlength\abovecaptionskip{0pt}
	\setlength\belowcaptionskip{0pt}
	\centering
	\setlength\fwidth{\columnwidth}
	\setlength\fheight{.25\columnwidth}
	\input{figurestikz/drl_apps_int_over_ks_slice1_throughput}
	\caption{\glspl{ecdf} of throughput of the \gls{embb} slice for xApps $a_7$ and $a_8$, highlighting the expected \gls{kpm} conflict for that control parameter.}
	\label{fig:drl_apps_int_over_ks_slice1_throughput}
\end{figure}

\subsubsection{Stochastic xApps}\label{sec:conflict-detection-evaluation-deterministic-apps}
To highlight differences between xApps with conflicting intents, we consider xApps $a_1$-$a_5$ from Section~\ref{sec:xapps:deterministic}. We report allocated \glspl{prb} for all slices, while only downlink throughput and buffer size are shown for \gls{embb} and \gls{urllc}, respectively. Tab.~\ref{tab:distances_numbers_deterministic} reports conflict values for both \gls{embb} and \gls{urllc}, as well as performance variation when two xApps with different severity indexes (i.e., $a_1$, $a_2$ and $a_5$) execute at the same time.

Since xApps $a_1$ and $a_2$ have similar \glspl{ecdf} (see Fig.~\ref{fig:stepcurves_gaussians}), Tab.~\ref{tab:distances_numbers_deterministic} shows low \gls{int} distance for \gls{embb} and \gls{urllc} slices with respect to both direct (i.e., \gls{prb} number) and \gls{kpm} (i.e., throughput and buffer size) conflicts. On the contrary, $a_1$ and $a_5$ show high \gls{int} and \gls{ks} distances with respect to \gls{prb} number and throughput (i.e., high direct and \gls{kpm} conflicts). However, $a_1$ and $a_5$ have low \gls{kpm} conflict with respect to buffer size for \gls{urllc}, which has an \gls{int} distance of $0.01$. These differences are also illustrated in Fig.~\ref{fig:exp1_int_over_ks_slice1_throughput}, where we focus on conflicts related to throughput of the \gls{embb} slice.
This demonstrates the importance of individually analyzing direct and \gls{kpm} conflicts, as large direct conflicts (i.e., \gls{int} distance between $a_1$ and $a_5$ with respect to \gls{prb} number for \gls{urllc} slice) can result in low \gls{kpm} conflicts (i.e., \gls{urllc} buffer size in the same case) due, for example, to traffic demand.

\begin{table}[b!]
	\addtolength{\tabcolsep}{-2pt} %
	\setlength\abovecaptionskip{5pt}
	\setlength\belowcaptionskip{5pt}
	\centering
	\caption{Direct and \gls{kpm} conflict analysis taking $a_1$ as the reference xApp.}
	\begin{tabular}{ c *{2} c *{6}{S[table-number-alignment = center]} }
		\arrayrulecolor{black} %
		\toprule
		\textbf{Slice} & \textbf{Variable} & \multicolumn{3}{c}{\textbf{$a_1$-$a_2$}} & \multicolumn{3}{c}{\textbf{$a_1$-$a_5$}}                                                                                          \\
		\cmidrule(lr){3-5} \cmidrule(lr){6-8}
		               &                   & {$D^{\gls{ks}}_{1,2}$}                   & {$D^{\gls{int}}_{1,2}$}                  & {Variation [\%]} & {$D^{\gls{ks}}_{1,5}$} & {$D^{\gls{int}}_{1,5}$} & {Variation [\%]} \\
		\midrule
		\gls{embb}     & {\glspl{prb}}     & 0.47                                     & 0.29                                     & -  8.64          & 1.00                   & 0.81                    & - 25.93          \\
		\gls{embb}     & {Throughput}      & 0.13                                     & 0.13                                     & - 16.42          & 0.82                   & 0.41                    & - 31.80          \\
		\midrule
		\gls{urllc}    & {\glspl{prb}}     & 0.48                                     & 0.28                                     & + 51.37          & 1.00                   & 0.65                    & + 94.47          \\
		\gls{urllc}    & {Buffer Size}     & 0.07                                     & 0.01                                     & - 84.21          & 0.14                   & 0.01                    & - 76.53          \\
		\bottomrule
	\end{tabular}
	\label{tab:distances_numbers_deterministic}
	\vspace{-0.2cm}
\end{table}

\ifexttikz
	\tikzsetnextfilename{exp1_int_over_ks_slice1_throughput}
\fi
\begin{figure}[t!]
	\setlength\abovecaptionskip{0pt}
	\setlength\belowcaptionskip{0pt}
	\centering
	\setlength\fwidth{\columnwidth}
	\setlength\fheight{.25\columnwidth}
	\input{figurestikz/exp1_int_over_ks_slice1_throughput}
	\caption{\glspl{ecdf} of throughput of the \gls{embb} slice for xApps $a_1$, $a_2$, and $a_5$, highlighting the expected \gls{kpm} conflict for that control parameter.}
	\label{fig:exp1_int_over_ks_slice1_throughput}
\end{figure}

\subsubsection{Importance of Mitigating Conflicts and Impact on \kpms}\label{sec:exp-mitigation}
Mitigating conflicts is essential to avoid conflicting control decisions causing performance degradation and unstable behavior, as previously shown in Fig. \ref{fig:motivational}, where $a_1$ and $a_5$ were used as the xApp \gls{es} and \gls{tm}, respectively.

\name evaluates how two applications generate control policies that conflict
on action values (e.g., how far away the two actions are) and \glspl{kpm}
(e.g., how to actions impact the \glspl{kpm}).
Conflicts are differentiated into three categories to identify which
conflicts only cause different actions (but same \glspl{kpm}), and which
cause different actions with different \gls{kpm} values.
Therefore, the system provides you with statistical information on how two
applications will impact system performance, and then the operator can
determine a threshold to identify what is a severe conflcit that needs to be
mitigated/prevented, and what is a conflict that can be tolerated.

In Tab.~\ref{tab:distances_numbers_deterministic}, we associate \gls{ks} and \gls{int} distances with performance variation.
We consider $a_1$ as the baseline xApp and evaluate distance and \kpm values if compared to $a_2$ and $a_5$.
As shown in Fig.~\ref{fig:stepcurves_gaussians} and summarized in Tab.~\ref{tab:distances_numbers_deterministic}, $a_1$ and $a_2$ have a low direct and \gls{kpm} conflict with \gls{ks} and \gls{int} with respect to \gls{prb} number allocated to \gls{embb} of 0.29 and 0.13, respectively. If compared to $a_1$, Tab.~\ref{tab:distances_numbers_deterministic} shows that $a_2$ allocates $\approx$8.6\% less \glspl{prb} to \gls{embb} with an \gls{int} severity index $\sigma^{\mathrm{K}}_{1,2}=0.21$ (reported in Tab.~\ref{tab:distances_apps_a3_2_a7_embb_int_averaged}), which results in a $\approx$16.4\% reduction in throughput. On the contrary, $a_5$ has a \gls{ks} and \gls{int} distance of respectively 0.81 and 0.41, and an \gls{int} severity index $\sigma^{\mathrm{K}}_{1,5}=0.59$. This highlights a more severe conflict that is demonstrated by the fact that $a_5$ allocates $\approx$26\% less \glspl{prb} to \gls{embb} (i.e., $\approx$2.8$\times$ less than $a_2$), which results in a 31.8\% reduction in throughput (i.e., 1.92$\times$ higher than $a_2$).

However, Tab.~\ref{tab:distances_apps_a3_2_a7_urllc_int_averaged} shows that \gls{urllc} still enjoys low buffer size even in the case of xApps with high direct conflict (i.e., $a_1$ and $a_5$), and the severity of \gls{kpm} conflicts $\sigma^{\mathrm{K}}$ never exceeds 0.02. For example, Tab.~\ref{tab:distances_numbers_deterministic} shows that $a_2$ allocates to \gls{urllc} 51\% more \glspl{prb} than $a_1$, which results in a 84\% reduction in buffer size. That is, what the \gls{embb} slice perceives as a conflict, actually benefits \gls{urllc}, demonstrating the need for a fine-grained conflict analysis framework such as \name to evaluate how conflicts impact intents and target \kpms.

\begin{figure}[t!]
	\centering
	\setlength\fwidth{\columnwidth}
	\setlength\fheight{.25\columnwidth}
	\input{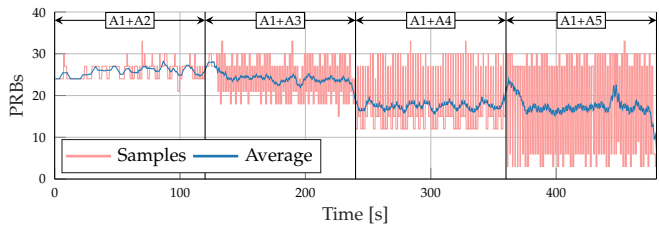}
	\vspace{-0.3cm}
	\caption{Impact of conflicts on the stability of the network for the \gls{embb} slice.}
	\label{fig:exp2_prb_embb_ts}
	\vspace{-0.3cm}
\end{figure}

Stability is another drawback caused by conflicts. In Fig.~\ref{fig:exp2_prb_embb_ts}, we report the \gls{prb} allocation for the \gls{embb} slice resulting from an 8-minute coexistence experiment in which we keep xApp $a_1$ always active, and iteratively activate a stochastic xApp every 2 minutes. We notice that the greater the direct conflict between the xApps (reported in Tab.~\ref{tab:combined_distances}), the larger the oscillations in the number of \glspl{prb} allocated to the \gls{embb} slice. This is also confirmed by the computation of three statistical metrics: namely the \gls{cov}, the \gls{stdev}, and the \gls{rmssd}. All three metrics considered represent a measure of the oscillation amplitude of the variable considered (assigned \glspl{prb} in this case), and they all grow as the conflict increases, as shown in Fig.~\ref{fig:exp2_prb_embb_covrmssd}. This might cause unstable behavior due to frequent updates of control parameters using conflicting policies that prevent a coordinated and orchestrated effort in satisfying intents.

\ifexttikz
	\tikzsetnextfilename{exp2_prb_embb_ts}
\fi

\ifexttikz
	\tikzsetnextfilename{exp2_prb_embb_covrmssd}
\fi
\begin{figure}[t!]
	\centering
	\setlength\fwidth{\columnwidth}
	\setlength\fheight{.25\columnwidth}
	\begin{tikzpicture}[
		auto,
		font=\tiny
	]

	\begin{axis}[%
			width=0.9\fwidth,
			height=0.9\fheight,
			at={(0\fwidth,-1.5\fheight)},
			scale only axis,
			xmin=0.5,
			xmax=4.5,
			xtick={1, 2, 3, 4},
			xticklabels={{$a_1$+$a_2$},{$a_1$+$a_3$},{$a_1$+$a_4$},{$a_1$+$a_5$}},
			xticklabel style={font=\scriptsize},
			xlabel style={font=\scriptsize\color{white!15!black}},
			xlabel={Phase},
			ymin=0,
			ymax=25,
			ytick={0, 5, 10, 15, 20, 25},
			ylabel style={font=\scriptsize\color{white!15!black}},
			ylabel={PRBs},
			axis background/.style={fill=white},
			title style={font=\bfseries},
			xmajorgrids,
			ymajorgrids,
			enlargelimits=false,
			legend style={inner sep=0pt, legend cell align=left,align=left,draw=white!15!black,anchor=north west,font=\scriptsize,at={(0.03,0.97)},cells={line width=0pt}},
			legend columns=1,
			reverse legend=false,
			ylabel shift=-3pt,
			xlabel shift=-3pt,
			clip mode=individual  %
		]
		\addplot [color=Paired-D, mark=asterisk, mark options={solid, Paired-D}]
		table[row sep=crcr]{%
				1 0.07541\\
				2 0.2013\\
				3 0.3929\\
				4 0.692\\
			};
		\addlegendentry{\gls{cov}}

		\addplot [color=Paired-B, mark=triangle, mark options={solid, Paired-B}]
		table[row sep=crcr]{%
				1 1.942\\
				2 4.882\\
				3 6.842\\
				4 12.04\\
			};
		\addlegendentry{\gls{stdev}}

		\addplot [color=Paired-F, mark=o, mark options={solid, Paired-F}]
		table[row sep=crcr]{%
				1 3.571\\
				2 9.457\\
				3 15.45\\
				4 23.95\\
			};
		\addlegendentry{\gls{rmssd}}

	\end{axis}

\end{tikzpicture}
	\vspace{-0.3cm}
	\caption{\gls{cov}, \gls{stdev}, and \gls{rmssd} for the \gls{embb} slice.}
	\label{fig:exp2_prb_embb_covrmssd}
	\vspace{-0.3cm}
\end{figure}

Figure~\ref{fig:bar_chart_comparison} shows the effectiveness of \name in evaluating the performance decrease when computing the severity index $\sigma^K$. The figure shows the trends of \glspl{kpm}, performance degradation, and severity index $\sigma^K$ for different pairs of application, with respect to the application $a_1$ running alone.
The further the xApp is from $a_1$ in terms of performance (here the downlink throughput is shown), the higher the severity index $\sigma^K$ is (for the sake of illustration, we show an example focusing on throughput only), and the higher is the performance degradation (computed as the relative drop in performance).

\ifexttikz
	\tikzsetnextfilename{bar_chart_comparison}
\fi
\begin{figure}[b!]
	\setlength\abovecaptionskip{0pt}
	\setlength\belowcaptionskip{0pt}
	\centering
	\setlength\fwidth{\columnwidth}
	\setlength\fheight{.7\columnwidth}
	\usepgfplotslibrary{patchplots,groupplots,colorbrewer}
\pgfplotsset{
	compat=newest,
	plot coordinates/math parser=false,
	every tick label/.style={font=\tiny},
	colormap/Paired
}

\begin{tikzpicture}
	\begin{axis}[
			width=0.9\fwidth,
			height=\fheight,
			bar width=0.2,        %
			ybar,
			ymin=0, ymax=5,
			axis y line*=left,
			axis x line=bottom,
			xmin=0.5, xmax=5.5,
			xtick={1,2,3,4,5},
			xticklabels={{$a_1$},{$a_1+a_2$},{$a_1+a_3$},{$a_1+a_4$},{$a_1+a_5$}},
			xlabel={Applications},
			ylabel={Throughput [Mbps]},
			ylabel style={color=black},
			ytick={1,2,3,4,5},
			yticklabel style={color=black},
			ymajorgrids,
			enlarge x limits=0.15,
		]
	\end{axis}

	\begin{axis}[
			width=0.9\fwidth,
			height=\fheight,
			bar width=0.2,        %
			ybar,
			ymin=0, ymax=1,
			axis y line*=right,
			axis x line=none,
			xmin=0.5, xmax=5.5,
			ylabel={Degradation/Severity Index []},
			ylabel style={color=black},
			yticklabel style={color=black},
			ytick={0,0.2,0.4,0.6,0.8,1.0},
			enlarge x limits=0.15,
			legend style={inner sep=2pt, legend cell align=left,align=left,
					draw=white!15!black,anchor=north east,font=\scriptsize,
					at={(0.99,0.99)},cells={line width=0pt}}
		]
		\addplot[ybar, bar shift=-0.25cm, fill=Paired-B, draw=black]
		table[
				row sep=crcr,
				x index=0,
				y expr=\thisrowno{1}/5
			] {
				1 4.193\\
				2 4.193\\
				3 4.123\\
				4 3.107\\
				5 2.536\\
			};
		\addlegendentry{Median Throughput}

		\addplot[ybar, bar shift=0cm, fill=Paired-F, draw=black]
		table[row sep=crcr] {
				1 0.0\\
				2 0.0\\
				3 0.017\\
				4 0.259\\
				5 0.395\\
			};
		\addlegendentry{Degradation}

		\addplot[ybar, bar shift=0.25cm, fill=Paired-D, draw=black]
		table[row sep=crcr] {
				1 0.000\\
				2 0.147\\
				3 0.243\\
				4 0.236\\
				5 0.461\\
			};
		\addlegendentry{Severity Index}
	\end{axis}
\end{tikzpicture}
	\caption{Bar chart with median throughput, the degradation (computed as the relative drop in performance), and the severity index $\sigma^K$ (computed as \gls{int} distance of throughput) of apps $a_1$ alone and app $a_1$ with $a_2$, $a_3$, $a_4$, or $a_5$.}
	\label{fig:bar_chart_comparison}
\end{figure}

Finally, in Tab.~\ref{tab:instantiation}, we provide two examples for different values of the conflict tolerance $\delta^\mathrm{TOL}$ with respect to the \gls{embb} slice and its conflict severity indexes from Tab.~\ref{tab:distances_apps_a3_2_a7_embb_int_averaged}. Specifically, we show how conflict tolerance specified by operators impacts the number of xApps that can coexist in the Near-RT \gls{ric}. We show the xApps that can be instantiated for a certain value of $\delta^\mathrm{TOL}$ in green, and those that cannot coexist due to high conflicts in red. We notice that low tolerance (i.e., $\delta^\mathrm{TOL}=0.25$) heavily limits the number of coexisting xApps, while a larger tolerance threshold (i.e., $\delta^\mathrm{TOL}=0.5$) leads to a more diverse xApp deployment.
Combining these results with the performance degradation reported in Tab.~\ref{tab:distances_numbers_deterministic}, we show that \name can effectively help operators in preventing deployment of applications that would impact \kpms above a certain tolerance threshold. For example, setting $\delta^{\mathrm{TOL}}\!=\!0.25$ limits coexistence of xApp $a_1$ to $a_2$ only (maximum decrease of throughput for the \gls{embb} slice of $-16.42\%$) and prevents, for instance, deployment of xApp $a_5$ which would reduce the same \kpm by approximately 32\%.

\begin{table}[t!]
	\centering
	\setlength\abovecaptionskip{0pt}
	\setlength\belowcaptionskip{-2pt}
	\caption{Coexistence under tolerance $\delta^{\mathrm{TOL}}\!=\!0.25$ (left) and $\delta^{\mathrm{TOL}}\!=\!0.5$ (right).} \label{tab:instantiation}
	\begin{tabular}{c|c}
		\begin{tabular}{c*{6}{|p{0.45cm}}}
			                          & \multicolumn{1}{c}{$a_1$}                                            & \multicolumn{1}{c}{$a_2$}                                            & \multicolumn{1}{c}{$a_3$}                                            & \multicolumn{1}{c}{$a_4$}                                            & \multicolumn{1}{c}{$a_5$}                                            \\
			\hhline{~*6{|-}|}
			$a_1$                     & \cellcolor{Paired-C!80} \textcolor{Paired-D!50!black}{0.00}          & \cellcolor{Paired-C!80} \textcolor{Paired-D!50!black}{0.21}          & \cellcolor{Paired-F!65!Paired-E} \textcolor{Paired-F!20!black}{0.35} & \cellcolor{Paired-F!65!Paired-E} \textcolor{Paired-F!20!black}{0.33} & \cellcolor{Paired-F!65!Paired-E} \textcolor{Paired-F!20!black}{0.59} \\
			\hhline{~*6{|-}|}
			\multicolumn{1}{c}{$a_2$} & \cellcolor{Paired-C!80} \textcolor{Paired-D!50!black}{0.21}          & \cellcolor{Paired-C!80} \textcolor{Paired-D!50!black}{0.00}          & \cellcolor{Paired-F!65!Paired-E} \textcolor{Paired-F!20!black}{0.28} & \cellcolor{Paired-F!65!Paired-E} \textcolor{Paired-F!20!black}{0.26} & \cellcolor{Paired-F!65!Paired-E} \textcolor{Paired-F!20!black}{0.55} \\
			\hhline{~*6{|-}|}
			\multicolumn{1}{c}{$a_3$} & \cellcolor{Paired-F!65!Paired-E} \textcolor{Paired-F!20!black}{0.35} & \cellcolor{Paired-F!65!Paired-E} \textcolor{Paired-F!20!black}{0.28} & \cellcolor{Paired-C!80} \textcolor{Paired-D!50!black}{0.00}          & \cellcolor{Paired-C!80} \textcolor{Paired-D!50!black}{0.19}          & \cellcolor{Paired-F!65!Paired-E} \textcolor{Paired-F!20!black}{0.48} \\
			\hhline{~*6{|-}|}
			\multicolumn{1}{c}{$a_4$} & \cellcolor{Paired-F!65!Paired-E} \textcolor{Paired-F!20!black}{0.33} & \cellcolor{Paired-F!65!Paired-E} \textcolor{Paired-F!20!black}{0.26} & \cellcolor{Paired-C!80} \textcolor{Paired-D!50!black}{0.19}          & \cellcolor{Paired-C!80} \textcolor{Paired-D!50!black}{0.00}          & \cellcolor{Paired-C!80} \textcolor{Paired-D!50!black}{0.00}          \\
			\hhline{~*6{|-}|}
			\multicolumn{1}{c}{$a_5$} & \cellcolor{Paired-F!65!Paired-E} \textcolor{Paired-F!20!black}{0.59} & \cellcolor{Paired-F!65!Paired-E} \textcolor{Paired-F!20!black}{0.55} & \cellcolor{Paired-F!65!Paired-E} \textcolor{Paired-F!20!black}{0.48} & \cellcolor{Paired-C!80} \textcolor{Paired-D!50!black}{0.00}          & \cellcolor{Paired-C!80} \textcolor{Paired-D!50!black}{0.00}          \\
			\hhline{~*6{|-}|}
		\end{tabular}
		 & \hspace{0.4cm}
		\begin{tabular}{c*{6}{|p{0.45cm}}}
			                          & \multicolumn{1}{c}{$a_1$}                                            & \multicolumn{1}{c}{$a_2$}                                            & \multicolumn{1}{c}{$a_3$}                                   & \multicolumn{1}{c}{$a_4$}                                   & \multicolumn{1}{c}{$a_5$}                                            \\
			\hhline{~*6{|-}|}
			$a_1$                     & \cellcolor{Paired-C!80} \textcolor{Paired-D!50!black}{0.00}          & \cellcolor{Paired-C!80} \textcolor{Paired-D!50!black}{0.21}          & \cellcolor{Paired-C!80} \textcolor{Paired-D!50!black}{0.35} & \cellcolor{Paired-C!80} \textcolor{Paired-D!50!black}{0.33} & \cellcolor{Paired-F!65!Paired-E} \textcolor{Paired-F!20!black}{0.59} \\
			\hhline{~*6{|-}|}
			\multicolumn{1}{c}{$a_2$} & \cellcolor{Paired-C!80} \textcolor{Paired-D!50!black}{0.21}          & \cellcolor{Paired-C!80} \textcolor{Paired-D!50!black}{0.00}          & \cellcolor{Paired-C!80} \textcolor{Paired-D!50!black}{0.28} & \cellcolor{Paired-C!80} \textcolor{Paired-D!50!black}{0.26} & \cellcolor{Paired-F!65!Paired-E} \textcolor{Paired-F!20!black}{0.55} \\
			\hhline{~*6{|-}|}
			\multicolumn{1}{c}{$a_3$} & \cellcolor{Paired-C!80} \textcolor{Paired-D!50!black}{0.35}          & \cellcolor{Paired-C!80} \textcolor{Paired-D!50!black}{0.28}          & \cellcolor{Paired-C!80} \textcolor{Paired-D!50!black}{0.00} & \cellcolor{Paired-C!80} \textcolor{Paired-D!50!black}{0.19} & \cellcolor{Paired-C!80} \textcolor{Paired-D!50!black}{0.48}          \\
			\hhline{~*6{|-}|}
			\multicolumn{1}{c}{$a_4$} & \cellcolor{Paired-C!80} \textcolor{Paired-D!50!black}{0.33}          & \cellcolor{Paired-C!80} \textcolor{Paired-D!50!black}{0.26}          & \cellcolor{Paired-C!80} \textcolor{Paired-D!50!black}{0.19} & \cellcolor{Paired-C!80} \textcolor{Paired-D!50!black}{0.00} & \cellcolor{Paired-C!80} \textcolor{Paired-D!50!black}{0.00}          \\
			\hhline{~*6{|-}|}
			\multicolumn{1}{c}{$a_5$} & \cellcolor{Paired-F!65!Paired-E} \textcolor{Paired-F!20!black}{0.59} & \cellcolor{Paired-F!65!Paired-E} \textcolor{Paired-F!20!black}{0.55} & \cellcolor{Paired-C!80} \textcolor{Paired-D!50!black}{0.48} & \cellcolor{Paired-C!80} \textcolor{Paired-D!50!black}{0.00} & \cellcolor{Paired-C!80} \textcolor{Paired-D!50!black}{0.00}          \\
			\hhline{~*6{|-}|}
		\end{tabular}
	\end{tabular}
\end{table}

\section{Conclusions} \label{sec:conclusions}

In this paper, we proposed \name, a framework to detect, characterize, and mitigate conflicts in the \oran ecosystem. \name leverages statistical information on \oran applications and a set of hierarchical graphs to determine the likelihood of conflict emergence and the corresponding severity. We derived a formal, data-driven model that \name uses to capture the impact of conflicts on control parameters and target \kpms, thus providing useful insights on how
conflicts affect the network performance and the operator's intents. We also proposed a tunable conflict mitigation strategy that uses \name statistical analysis to determine which \oran applications can coexist, and which should not be deployed to prevent performance degradation.
We prototyped \name on a real-world \oran testbed and carried out an experimental campaign that demonstrated \name's effectiveness in characterizing conflicts and providing insights for informed deployment decisions.

Future work will focus on three aspects: (i) extracting a model to capture the relationship between conflict severity and the performance degradation of the \ran to predict performance degradation before applications are deployed; (ii) how the impact of asynchronous actions generated by different applications affects conflicts severity; and (iii) what strategies can be developed and implemented by the network operator to adaptively express thresholds for the mitigation phase.

\bibliographystyle{IEEEtran}
\bibliography{bibliography}

\begin{IEEEbiography}[{\includegraphics[width=1in,height=1.25in,clip,keepaspectratio]{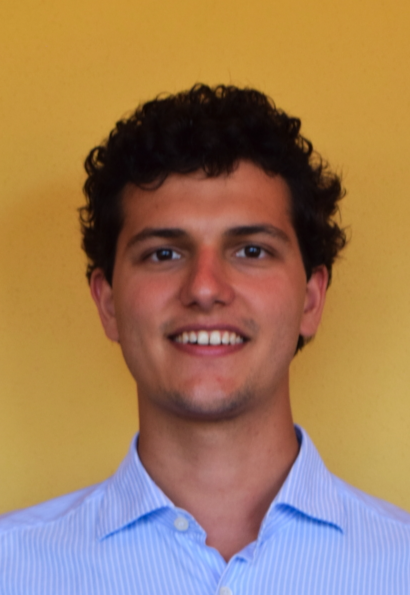}}]{Pietro Brach del Prever} is a Ph.D.\ Candidate at the Institute for the Wireless Internet of Things at Northeastern University. He graduated at Politecnico di Torino in Mechatronic Engineering in 2022, and in Mechanical Engineering in 2020. He visited the University of Porto in 2019 and the Technion Israel Institute of Technology in 2021 as an exchange student. He also graduated from Alta Scuola Politecnica, a joint honor program of Politecnico di Torino and Politecnico di Milano for the top 1\% students of both universities. His research interests include wireless networks optimization, and intrabody communication.
\end{IEEEbiography}

\vspace{-0.5cm}

\begin{IEEEbiography}[{\includegraphics[width=1in,height=1.25in,clip,keepaspectratio]{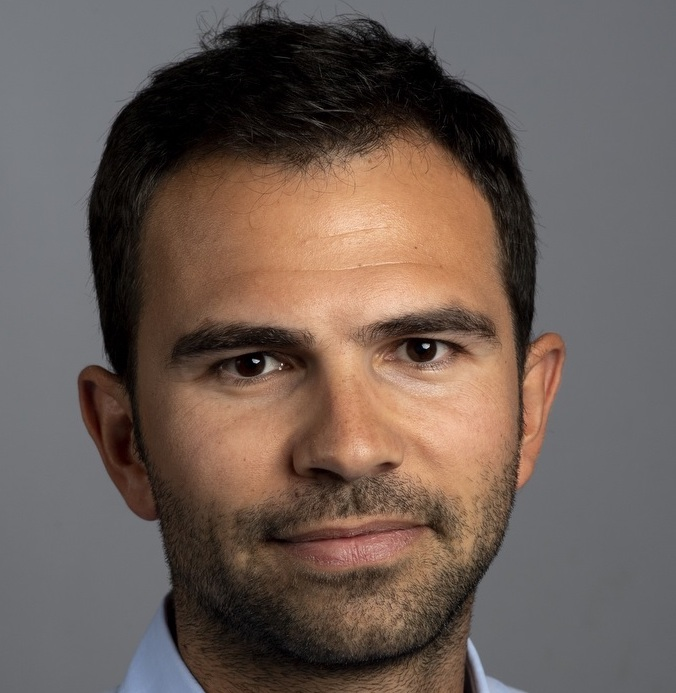}}]{Salvatore D'Oro} is a Research Associate Professor at Northeastern University. He received
	his Ph.D. degree from the University of Catania
	and is an area editor of Elsevier
	Computer Communications journal. He serves
	on the TPC of
	IEEE INFOCOM, IEEE CCNC \& ICC and
	IFIP Networking. He is one of the contributors to
	OpenRAN Gym, the first open-source research
	platform for AI/ML applications in the Open RAN.
	His research interests include optimization, AI \& network slicing for NextG Open RANs.
\end{IEEEbiography}

\vspace{-0.5cm}

\begin{IEEEbiography}[{\includegraphics[width=1in,height=1.25in,clip,keepaspectratio]{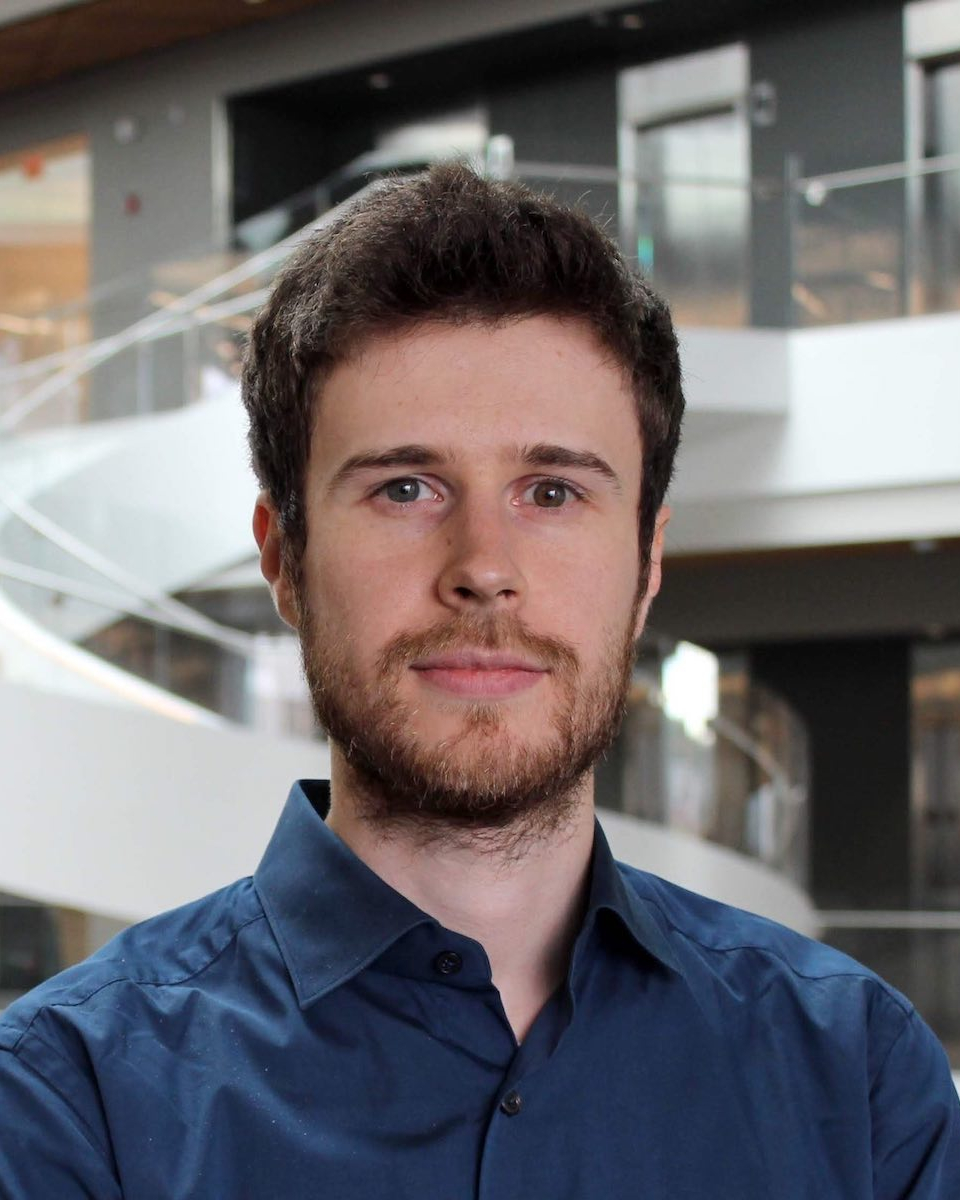}}]{Leonardo Bonati}
	is an Associate Research Scientist at the Institute for the Wireless Internet of Things, Northeastern University. He received a Ph.D.\ degree in Computer Engineering from Northeastern University in 2022. His research focuses on softwarized approaches for the Open RAN of next generation of cellular networks, on O-RAN-managed networks, and on network automation and orchestration. He served as guest editor of the special issue of Elsevier's Computer Networks journal on Advances in Experimental Wireless Platforms and Systems.
\end{IEEEbiography}

\vspace{-0.5cm}

\begin{IEEEbiography}[{\includegraphics[width=1in,height=1.25in,clip,keepaspectratio]{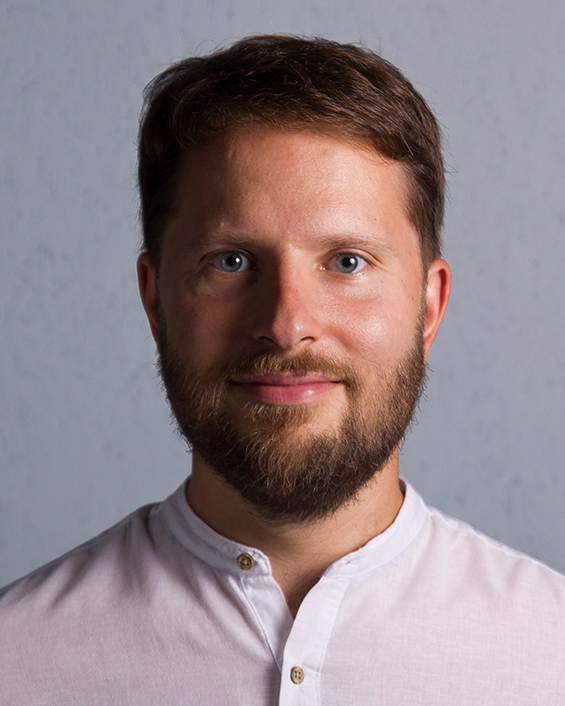}}]{Michele Polese} is a Research Assistant Professor at Northeastern University. He received his Ph.D. from the University of Padova, while his research is funded by the US NSF, OUDS, and NTIA. He holds several best paper awards and the '22 Mario Gerla Award for Research in Computer Science. He is a TPC co-chair for WNS3 '22, an Associate Technical Editor for the IEEE Communications Magazine, and a Guest Editor in JSAC's Special Issue on Open RAN.
\end{IEEEbiography}

\vspace{-0.5cm}

\begin{IEEEbiography}[{\includegraphics[width=1in,height=1.25in,clip,keepaspectratio]{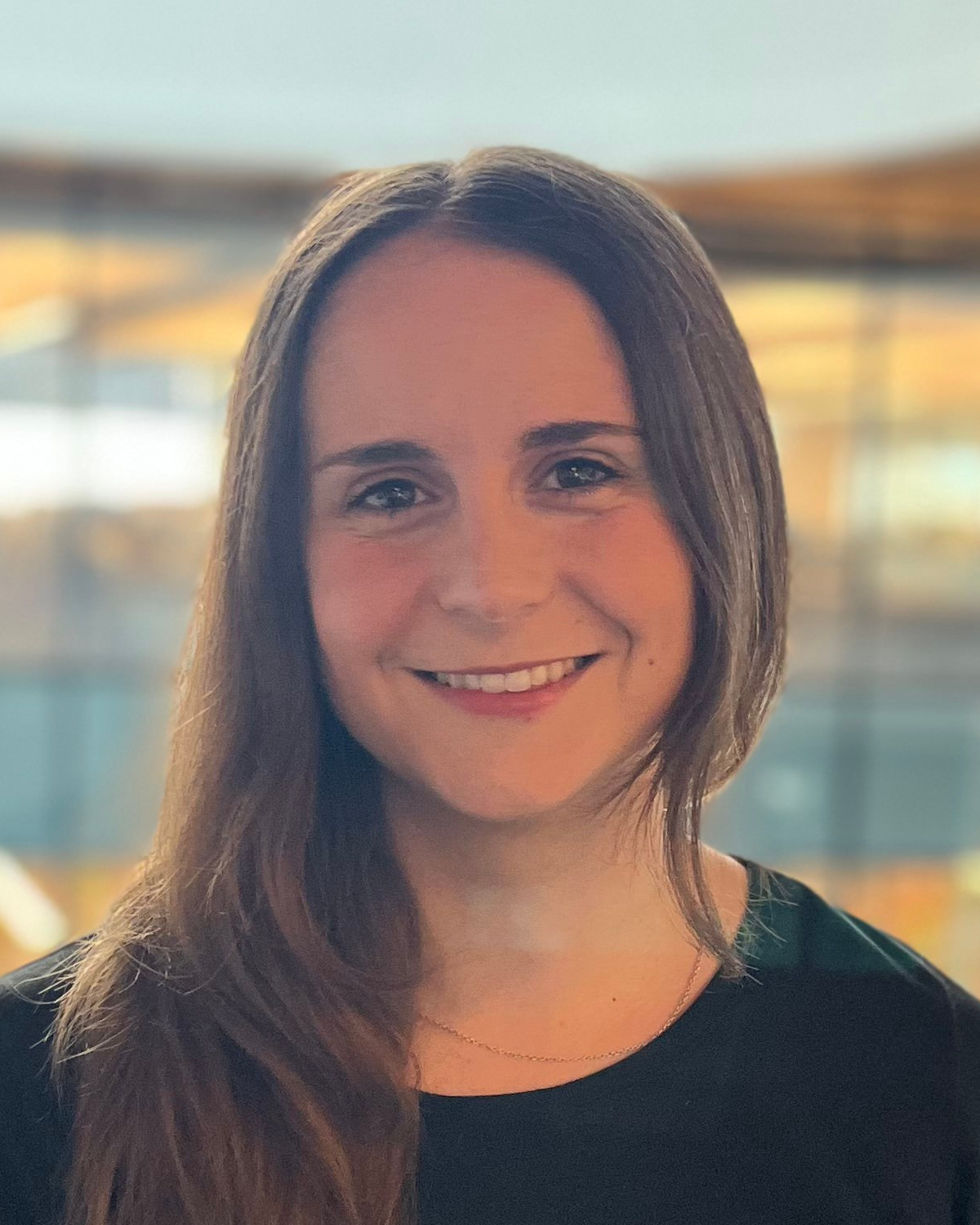}}]{Maria Tsampazi}
	received her MEng Degree in ECE from National Technical University of Athens, Greece in '21. She is a Ph.D. Candidate in Electrical Engineering at the Institute for the Wireless Internet of Things, NEU. Her research focuses on NextG networks and intelligent resource allocation in Open RAN. She has received academic awards sponsored by the US NSF, IEEE ComSoc \& NEU, and is a National Spectrum Consortium Woman in Spectrum scholarship recipient. Maria has collaborated with government and industry organizations such as the US DoT and Dell Technologies.
\end{IEEEbiography}

\vspace{-0.5cm}

\begin{IEEEbiography}[{\includegraphics[width=1in,height=1.25in,clip,keepaspectratio]{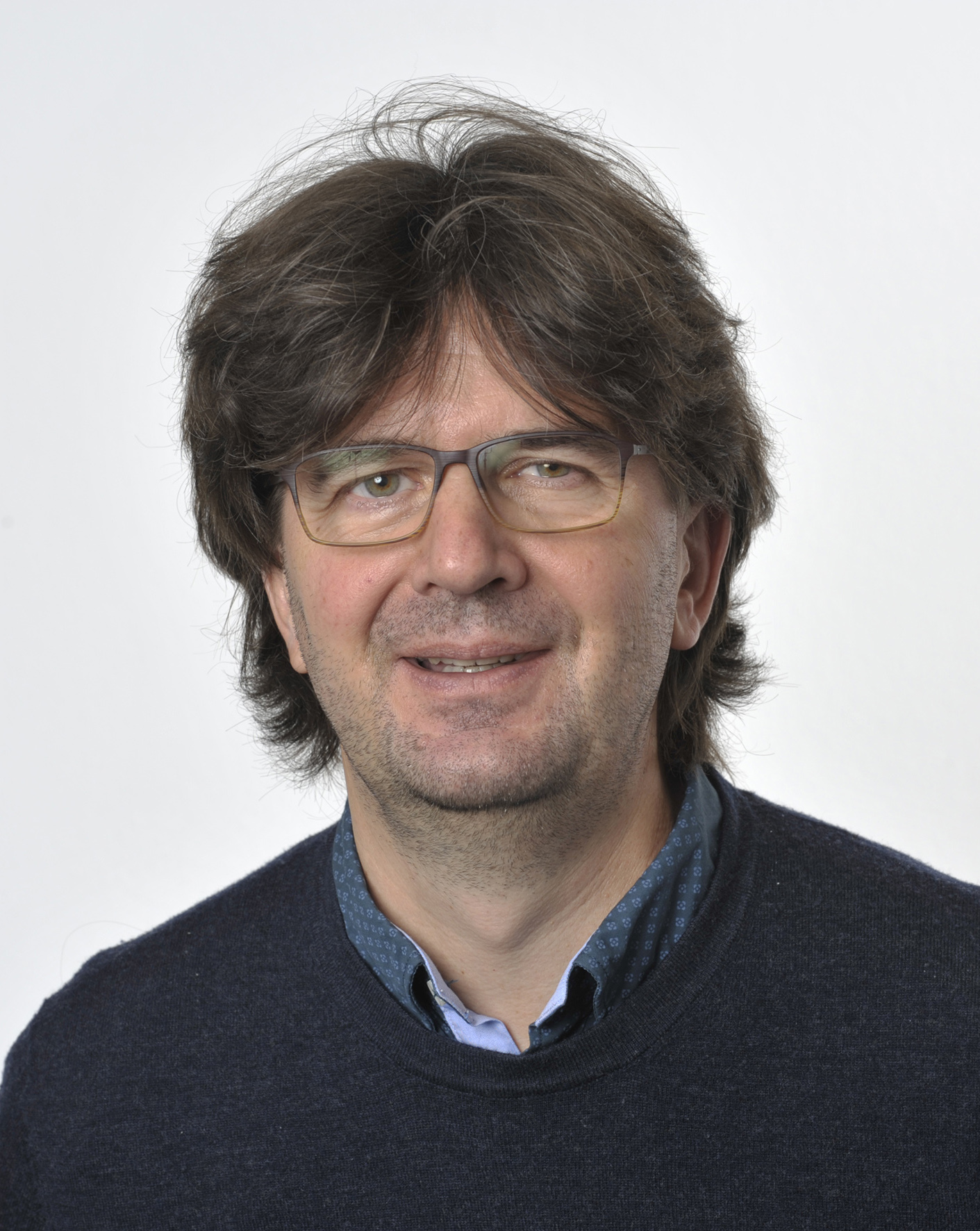}}]{Heiko Lehmann}
	Heiko Lehmann is Tribe Lead for Cybersecurity and Digital Twin at in Deutsche Telekom's Group Technology Division. He is affiliated with T-Labs. Trained a physicist, he received a Ph.D.\ in Theoretical Physics from Humboldt University Berlin in 1992. Following postdoctoral academic work at Oxford University and the German National Society for Mathematics and Informatics, he joined the telematics subsidiary of Volkswagen Group in 1999 where he held the position of Innovation Manager. In 2006, Lehmann joined T-Labs where he took over responsibility for a portfolio of innovation projects. Recently, he is focusing on Artificial Intelligence and its application to Digital Twins, cybersecurity, self-organization, complex systems and optimization in a wide variety of application areas.
	Lehmann has published extensively in theoretical physics, informatics, engineering and business topics.
\end{IEEEbiography}

\vspace{-0.5cm}

\begin{IEEEbiography}[{\includegraphics[width=1in,height=1.25in,clip,keepaspectratio]{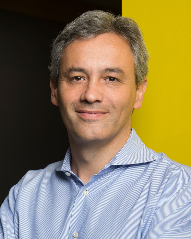}}]{Tommaso Melodia}
	is the William Lincoln Smith Professor at Northeastern
	University,
	Director of the Institute for the Wireless Internet
	of Things and Director of Research for the Platforms for Advanced Wireless Research, a $100M$ public-private partnership to advance the US wireless
	ecosystem. He received his Ph.D. from the
	Georgia Institute of Technology and is
	a recipient of the NSF
	CAREER award. He has served as
	Associate Editor of IEEE Transactions on Wireless Communications, Transactions on Mobile Computing and Elsevier Computer Networks, and as TPC Chair
	for IEEE INFOCOM '18, General Chair for IEEE SECON '19, ACM
	Nanocom '19, and WUWnet '14.  His research on the experimental evaluation of IoT and wireless
	networked systems has been funded by the NSF, the Air Force \& Army Research Laboratories, the Office of Naval Research, and
	DARPA. He is an IEEE Fellow and ACM Senior Member.
\end{IEEEbiography}

\vfill

\end{document}